\documentclass[a4paper,12pt]{article}
\pdfoutput=1 

\usepackage[utf8]{inputenc}
\usepackage[english]{babel}
\usepackage{amssymb, amsmath}
\usepackage{amsthm}
\usepackage{amsfonts}
\usepackage{setspace}
\usepackage{graphicx}
\usepackage{slashed}
\usepackage[sorting=none, backend=bibtex]{biblatex}
\bibliography{ref}

\usepackage{array,booktabs}
\usepackage{lmodern}
\usepackage{tikz}
\usepackage{microtype}
\usepackage{hyperref}
\usepackage{mathtools}
\usetikzlibrary{shapes.gates.logic.US, decorations.markings,decorations.pathreplacing,calc}

\newcommand{\p}{\partial}
\newcommand{\beq}{\begin{equation}}
\newcommand{\eeq}{\end{equation}}
\newcommand{\eeeem}{\end{multline}}
\newcommand{\bem}{\begin{multline}}
\newcommand{\bqa} {\begin{eqnarray}}
\newcommand{\eqa} {\end{eqnarray}}
\newcommand{\eps}{\varepsilon}
\newcommand{\ep}{\epsilon}
\newcommand{\bmul}{\begin{multline}}
\newcommand{\emul}{\end{multline}}
\DeclareMathOperator{\Tr}{Tr}

\DeclareMathOperator{\arccosh}{arccosh}

\DeclareMathOperator{\Type}{Type}
\DeclareMathOperator{\Fl}{Fl}

\def \pr {\partial}
\def \ra {\rightarrow}

\renewcommand\Im{\operatorname{Im}}

\def \nsusy {\mathcal{N}}
\def \Rbb {\mathbb{R}}

\def \l {\left(}
\def \r {\right)}

\def \lal {\langle}
\def \rr {\rangle}

\def \la {\lambda}
\def \La {\Lambda}

\def \ep {\epsilon}

\begin{document}
\begin{titlepage}

\begin{center}
{ \Large \bf Bands and gaps in Nekrasov partition function}
\end{center}
\vspace{1mm}

\begin{center}

{\large

  A.~Gorsky$^{\,2,3}$,    A.~Milekhin$^{\,1,3,4}$\ and N.~Sopenko$^{\,1,2,3}$\ }

\vspace{3mm}

$^1$ Institute of Theoretical and Experimental Physics, B.Cheryomushkinskaya 25, Moscow 117218, Russia \\
$^2$ Moscow Institute of Physics and Technology, Dolgoprudny 141700, Russia \\
$^3$ Institute for Information Transmission Problems of Russian Academy of Science, B. Karetnyi 19, Moscow 127051, Russia\\
$^4$ Department of Physics, Princeton University, Princeton, NJ 08544, USA \\

\vspace{1cm} 

gorsky@itep.ru, milekhin@itep.ru, niksopenko@gmail.com

\end{center}

\vspace{1cm}

\begin{center}

{\large \bf Abstract}

\end{center}

We discuss the effective twisted superpotentials of 2d $\mathcal{N}=(2,2)$ theories arising upon the reduction of 4d $\mathcal{N}=2$ gauge theories on the $\Omega$-deformed 
cigar-like geometry. We explain field-theoretic origins of the gaps in the spectrum in the corresponding quantum mechanical (QM) systems. 
We find local 2d descriptions of the physics near these gaps by resumming the non-perturbative part of the twisted superpotential and discuss 
arising wall-crossing phenomena. The interpretation of the associated phenomena in the classical Liouville theory and in the scattering of two heavy states in $AdS_3$ 
gravity is suggested. Some comments concerning a possible interpretation of the band structure in QM in terms of the Schwinger monopole-pair production in 4d are presented.

\end{titlepage}

\tableofcontents

\section{Introduction}

The Seiberg-Witten (SW) solution of $\mathcal{N}=2$ SYM theories \cite{SW} provides an 
impressive example of how the summation over the non-perturbative configurations
in the non-integrable quantum field theory amounts for the powerful information concerning the low-energy effective action and the spectrum of the stable particles. The combination of the educated
physical guesses and the analytic tools allowed to treat the strong coupling regime 
exactly. In the case of pure $SU(2)$ theory it was discovered that the naive classical singularity where the $W$-boson becomes massless  split into 
two singular points, where the monopole and the dyon become massless instead, enclosed by the curve of marginal stability (CMS) within which the W-boson becomes unstable and no longer exists in the spectrum. If $\mathcal{N}=2$ SUSY is
softly broken down to $\mathcal{N}=1$ SUSY the singular points of the Coulomb moduli space correspond to the vacuum states of the $\mathcal{N}=1$ SYM theory. 

The Seiberg-Witten result was reproduced microscopically by Nekrasov \cite{nekrasov, NO02} by considering the theory in the $\Omega$-background $\mathbf{C}^2_{\ep_1, \ep_2}$ that provides the IR regularization by localizing the integrals over the instanton moduli spaces on a set of isolated fixed points. 
The partition function is expressed as a sum of the contributions from each of these points. The Seiberg-Witten prepotential, which defines the low-energy effective
action, can be extracted from the
Nekrasov partition function as the leading term of the logarithm of the partition function in the limit $\epsilon_{1}, \epsilon_{2} \rightarrow 0$.

The information contained in the Seiberg-Witten solution can be usefully packed in the pair
of integrable Hamiltonian systems. First, the Seiberg-Witten curves  which encode the information
about the low-energy actions were identified with the spectral curves of the classical holomorphic integrable 
many-body systems \cite{gkmmm,warner,DW95,gmmm,fb1,fb2}. The type of the integrable many-body system
is in one-to-one correspondence with the matter content of the $\mathcal{N}=2$ SYM theory. The SW variables $a,a_D$ were
identified with the action and the dual action in the corresponding integrable system. The prepotential
on the other hand was naturally identified in terms of the second Whitham integrable system \cite{gkmmm,gmmm2}
that is the semiclassical limit of the (1+1) dimensional KdV or Toda field theory. In the Whitham theory
the variables $a,a_D$ form the symplectic pair while the prepotential $\mathcal{F}$ plays the role 
of the action variable which can be immediately seen from the 
relation $a_D=\frac{\partial \mathcal{F}}{\partial a}$
which is the analogue of the relation $p=\p S/ \p q$ in the Hamiltonian mechanics. The time in the Whitham
dynamics was identified with the logarithm of the instanton counting parameter $\log \Lambda$. The coordinates in the first integrable system were
identified with the positions of the defect branes in some internal coordinate \cite{fb1,fb2}.

The next step in this correspondence was made in \cite{NS09p1,NS09p2,NS09p3}. It was argued that by turning on an $\Omega$-background with only one non-zero parameter $\ep_1$ the many-body type integrable system gets quantized with a Planck constant $\hbar = \ep_1$. The Yang-Yang function (YY) of this system, which provides the Bethe Ansatz equations for a particular quantization, coincides with the effective twisted superpotential of the effective two-dimensional $\mathcal{N}=(2,2)$ theory that appears upon the reduction on the plane with $\Omega$-background turned on with the appropriate choice of the boundary conditions at the boundary of this plane, while the quantum states are identified with the vacua of this theory. The effective twisted superpotential arises as the leading term in the logarithm of the partition function in the Nekrasov-Shatashvili (NS) limit $\ep = \ep_2 \to 0$. A more precise description of this two dimensional theory was given in \cite{nw10}.

An interesting feature that appears in quantum many-body integrable systems is the presence of exponentially small in $\hbar$ gaps between the energy bands for a particular type of quantization.
 For example, for pure $U(2)$ gauge theory one of the quantizations leads to the famous Mathieu spectral problem which gives the spectrum of a particle in a periodic cosine potential.
 This gap between the bands has a transparent physical interpretation in this quantum mechanical problem and appears due to quantum reflection over the barrier. 
However, it looks rather mysterious from the gauge theoretic point of view. The four-dimensional $W$-boson leads to an infinite tower of  two-dimensional particles in
 the effective theory upon the reduction, and a quick inspection shows that some of this particles naively become massless precisely at those values of the Coulomb parameters,
 for which the splitting phenomenon appears, undermining the validity of the low-energy effective description. These special values of the Coulomb parameters also appear
 as poles in the instanton part of the superpotential. 
For each splitting point there are
infinitely many poles of all orders in the instanton series.

The goal of the present paper is to describe the physics near this loci. What we have found is that after a proper resummation of the non-perturbative part and after combining it 
with the perturbative contribution the naive pole singularities in the superpotential transform into the cuts. The fate of the corresponding two-dimensional $W$-boson mode is rather similar to the fate of the four-dimensional $W$-boson in the undeformed theory: it decays into lighter particles before becoming massless on the curve of marginal stability that encloses the cut. In our case this lighter particles are solitons which interpolate between the vacua corresponding to split quantum states.

Let us note that the procedure of trans-series resummation that we actually use is known in the mathematical literature
as the trans-asymptotic matching \cite{costin}. It concerns the reordering of the summation of trans-series
near the naive poles. The procedure
has been successfully applied for the Painleve I and Painleve II equations \cite{costin} and in the 
physical context it was applied for the trans-series for the gradient expansion of the 
hydrodynamical equations in \cite{dunnehydro}. It was argued that the leading term of the 
resummed trans-series obeys a nonlinear equation which is universal while 
the next terms of the expansion of the resummed trans-series can be derived via the
recursion relation from the leading term. We shall focus in this study on the 
leading term in the resummed trans-series for the twisted superpotential which 
locally describes the low-energy physics near the cuts when $\hbar$ is large.

Remarkably, the twisted superpotential for this local description has a rather familiar form. 
For pure $U(2)$ gauge theory it coincides with the superpotential of the sigma model on $\mathbf{CP}^1$ 
with the twisted mass which is proportional to the "distance" $\sim(2a-\hbar)$ from the naive 
singularities at $2a = \hbar$, and the two vacua of this sigma model correspond to split levels. 
This identification allows us to read out the BPS spectrum of light particles of the effective theory near the cuts, since the BPS spectrum for $\mathbf{CP}^1$-model was derived long time 
ago \cite{dorey98,shifman}. We also make a non-trivial check that in the local 
description of $U(N)$ gauge theory one has the sigma model on a flag variety, with the flag
type determined by the level splitting pattern.


The AGT correspondence \cite{agt} relates the Nekrasov partition function with the conformal blocks in the Liouville theory. The NS limit in the gauge theory corresponds to the
classical limit on the Liouville side. Hence it is natural to ask  what kind of 
non-perturbative phenomena in Liouville theory 
the non-perturbative phenomena  in QM correspond to. It is known that  (see, for instance \cite{piatek1,piatek2}) 
that the twisted superpotential of
$SU(2)$ SYM with adjoint matter coincides with the classical one-point
torus conformal block in the Liouville theory while $SU(2)$ SYM with fundamental matter coincides 
with the classical 4-point spherical conformal block. 
It is useful to map the classical torus 1-point conformal block 
to the particular 4-point spherical conformal block \cite{Fateev1,Poghossian10} hence all $SU(2)$ 
examples can be considered on equal footing.

The 2-point torus classical conformal block 
involving the $\Psi_{2,1}$ degenerate operator obeys the
Lame equations \cite{Fateev}. The Mathieu equation can be derived in a proper limit from Lame equation and its solution 
was identified with  the irregular conformal block involving coherent Gaiotto state \cite{gaiotto}.  Therefore we
shall argue that naive poles in intermediate dimensions in classical 4-point conformal block 
disappear upon the trans-series resummation.  

The classical Liouville conformal block  by $AdS_3/CFT_2$ 
holographic correspondence (see \cite{KrausLectures} for a review) is lifted to a particular process in  $AdS_3$ bulk gravity. 
It was identified with the on-shell action evaluated on the geodesic network in 3d gravity \cite{Fit1,asplund,kraus}.
The effects of operator insertions at the 
boundary  depend on the behavior of their dimensions in the classical limit. If the operator is heavy, that is its conformal
dimension is proportional to the central charge, it deforms the bulk and
amounts to BTZ black hole or conical defect depending on the value of the classical
conformal dimension. The light operators correspond to the geodesic motion
in the bulk perturbed by the heavy operators \cite{Fit1,Fit2,Fit3,belavin1,belavin2}. The accessory parameter in the Liouville theory
which is related to the energy in the  QM problems 
corresponds to the conserved Killing momentum on the gravity  side \cite{Fit1,asplund,kraus}.
Since we have explained
the mechanism of the disappearance of the naive poles in 
the classical Liouville conformal blocks, the natural
question concerns the meaning of the corresponding non-perturbative 
phenomena in  $AdS_3$ gravity. 

It turns
out that all four operators in the spherical conformal block are heavy 
and the intermediate dimension
is related to the heavy operator as well. The naive poles correspond
to the points where the intermediate state
becomes degenerate and in the limit of the 
large Planck constant(in QM sense) we are focusing on the "OPE limit"
in the conformal block. Since all operators are heavy we are dealing with the scattering  of the  degrees
of freedom which deform the bulk gravity. Not much is known about the scattering S-matrix
in $AdS_3$-gravity however some non-perturbative phenomena have been
considered for instance in \cite{krasnov1,krasnov2}. In particular it was shown  that there is
the exponentially suppressed process of the black hole formation
in collision of two particles representing conical defects \cite{krasnov2}.

Therefore the gravity problem at hand involves the extremal BTZ BH and
particles yielding conical singularities. Using the map between the 
classical conformal block and on-shall gravity action for the scattering
process we claim that the poles in the scattering amplitude as the
function of the intermediate dimension disappears upon the resummation
of trans-series and the CMS around the naive pole emerges. The CMS
can be considered as the curve on the plane of the complex intermediate
dimension  at fixed time or the curve at the complex time plane at 
fixed intermediate dimension. This suggests that probably the 
transition between early times and late times proceeds through
a kind of CMS. 

There is an interesting similarity of our case with the low-energy 
monopole scattering  \cite{AH85}. That process can be described
as the geodesic motion at the 2-monopole moduli space which 
has a non-trivial Atiyah-Hitchin metric - the non-perturbative
generalization of the Taub--NUT metric. There are specific geodesics
on the monopole moduli space which result in the excitation of the
motion along $S^1$ direction of the monopole moduli space. Physically
it means that the monopoles become dyons in the scattering process.
This happens because the initial angular momentum
of the monopoles gets transformed into the angular momentum 
of the electromagnetic field in the collision process. In the
physical space the process can be described by the exchange 
of the light $W$-boson between two heavy monopoles.

The $AdS_3$ gravity can be described via $SL(2,\mathbb{R})
\times SL(2,\mathbb{R})$ Chern-Simons theory \cite{townsend,witten88}. Hence 
we are dealing with the moduli space of flat connections and the
corresponding Teichmuller spaces(which are analogues of the
monopole moduli space for the monopole scattering). The 
twisted superpotential is the action for the Whitham dynamics
which on the other hand corresponds to the geodesic 
motion on the Teichmuller space similar to the monopole case.
It is natural to conjecture that similar to the monopole case 
the motion along the compact dimension of the moduli space
can be excited in the process.


It was conjectured in \cite{DB15} that some kind of 
the Schwinger-type phenomena takes place in SYM which
corresponds to the bounce-induced phenomena in QM however
no mechanism has been found. We shall present some arguments 
that the non-perturbative monopole pair
production in the external graviphoton field is relevant for the 
bounce induced phenomena in the QM. We shall also comment
on the relation of the Schwinger-like picture to the representation 
of the Nekrasov partition
function via the Myers effect in the external graviphoton field.


There were several previous studies addressing the issue of singularities in the effective twisted superpotential. In \cite{Becc}  the resummation procedure that we use in this paper was performed for finite-gap $\mathcal{N}=2^*$ $U(2)$ gauge theory. In \cite{Jeong17} the partition functions of half-BPS surface defects near the corresponding locus were studied. It is necessary
to mention \cite{Lukyanov} where the Mathieu equation has been obtained in the
mini-superspace approximation in the sine-Gordon model on a cylinder. The absence of the naive singularity in the plane of quasi-momentum in the sine-Gordon model
has clear a physical interpretation. The quasi-momentum is related to the 
effective constant electric field in the fermionic representation and the 
emergence of the cut instead of the logarithmic singularities is associated with the Schwinger-type pair production in the external electric field. 
\\

The paper is organized as follows. In section 2 we review some facts concerning the
Nekrasov partition function and Bethe/gauge correspondence. In section 3 we discuss the low-energy description of the effective $\mathcal{N}=(2,2)$ two-dimensional theory. We perform  the resummation of  non-perturbative contribution and show that the naive singularities of the superpotential in the Coulomb parameters get transformed into the cuts. We extract the local superpotentials near these cuts and discuss their physical implications. 
In section 4 the interpretation of this phenomena  in 
the classical Liouville theory and its holographic dual
is discussed. Some conjectures concerning the possible relation with the monopole-pair
production in the $\Omega$-background are presented in section 5. The considerations in sections 4 and 5 
are more qualitative however they suggest useful physical insights and analogies. The results and the open questions are collected in Conclusion.


\section{Generalities on the Bethe/gauge correspondence}
In this section we review the Bethe/gauge correspondence \cite{MNS97, GS07, GS08, NS09p1, NS09p2, NS09p3} to set up the conventions and formulate the problem. Throughout the paper we only discuss the case of theories with a single $U(N)$ gauge group.

\subsection{Instanton partition function}
First, let us recall the definition of the Nekrasov partition function \cite{nekrasov, NO02} of a four dimensional gauge theory in the $\Omega$-background $\mathbf{C}^2_{\ep, \hbar}$. 

Let $\mathbf{a}$ denote a set of complex scalars which parametrize the moduli space of vacua, $\mathbf{m}$ is set of masses for fundamental matter multiplets and $\Lambda^{2N} = \exp (2 \pi i \tau)$ is a generated mass scale that counts instantons. 

The full Nekrasov partition function consists of perturbative and non-perturbative contributions
\beq
Z(\mathbf{a},\mathbf{m},\Lambda; \ep, \hbar) = Z^{pert.}(\mathbf{a},\mathbf{m},\Lambda; \ep, \hbar) \times Z^{inst.}(\mathbf{a},\mathbf{m},\Lambda; \ep, \hbar)
\eeq
The non-perturbative part of the partition function is obtained by the equivariant localization on the instanton moduli space and is defined as follows. Let
\beq
\mathcal{V}_{\lambda} = \sum_{i=1}^{N} \sum_{(r,s) \in \lambda_{i}} e^{a_i + (r-1)\ep + (s-1) \hbar}, \,\,\,\,\,\,\, \mathcal{W} = \sum_{i=1}^{N} e^{a_i}, \,\,\,\,\, \mathcal{M} = \sum_{i=1}^{N_f} e^{m_i}
\eeq
\beq
\mathcal{T}_{\lambda} = - \mathcal{M} \mathcal{V}_{\lambda}^* + \mathcal{W} \mathcal{V}_{\lambda}^* + e^{\ep+\hbar} \mathcal{V}_{\lambda} \mathcal{W}^* - (1 - e^{\ep}) (1-e^{\hbar}) \mathcal{V}_{\lambda} \mathcal{V}_{\lambda}^*
\eeq
which appear as the characters of the natural bundles on the instanton moduli space at a fixed point, parametrized by a set $\{ \lambda_i \}$ of $N$ Young diagrams. The star-operation inverts all weights of a character: 
\beq
\biggl(\sum_{a} e^{w_a} \biggr)^* =  \sum_{a} e^{-w_a}
\eeq
The instanton partition function can be written as
\beq
Z^{inst.} (\mathbf{a} ,\mathbf{m}, \Lambda; \ep, \hbar) = \sum_{\{\lambda\}} \Lambda^{2N |\lambda|} e(\mathcal{T}_{\lambda})
\eeq
where
\beq
e(\sum_{a}e^{w_a}-\sum_b e^{w_b}) = \frac{\prod_b w_b}{\prod_a w_a}
\eeq
is a symbol that converts the sum of characters into the product of weights. 

The perturbative part is more subtle due to ambiguity of the boundary conditions at infinity \cite{bps/cft1}. It can be written as
\beq
Z^{pert.}(\mathbf{a},\mathbf{m},\Lambda;\ep,\hbar) = \Lambda^{-\frac{N}{\ep \hbar} \sum_{i=1}^{N} a_i^2} e\biggl( \frac{e^{\ep+\hbar}(\mathcal{M} \mathcal{W}^* -\mathcal{W} \mathcal{W}^*)}{(1-e^{\ep})(1-e^{\hbar}) }\biggr)
\eeq 
but since the character now has infinitely many terms a proper regularization is needed. Physically the first multiplier comes from the tree level contribution, while the first and the second terms in the character come from the one-loop contribution of the matter multiplets and $W$-bosons, respectively. 

Note that the perturbative contribution does depend on the instanton counting parameter since it defines the running coupling constant $\tau$.

\subsection{Nekrasov-Shatashvili limit}

The low-energy description of undeformed four dimensional gauge theory is characterized by the prepotential, which can be obtained as the limit of the deformed partition function
\beq
\mathcal{F}(\mathbf{a}, \mathbf{m}, \Lambda) =  \lim_{\ep,\hbar \to 0} \ep \, \hbar \log Z(\mathbf{a},\mathbf{m}, \Lambda; \ep,\hbar)
\eeq
It is known that it is related to some classical algebraic integrable system \cite{gkmmm, DW95}. In particular the moduli space of vacua of the theory coincides with the base of the Liouville fibration of this system. The underlying integrable system for a large class of quiver gauge theories was found in \cite{NP12}.

In \cite{NS09p3} a refinement of the correspondence with integrable systems was proposed. 
It was argued that the effective two dimensional theory, which appears in the limit $\ep \to 0$, is related to the corresponding quantized algebraic integrable system, and $\hbar$ plays the role of the quantization parameter. 

More precisely we can consider a four dimensional gauge theory on $\mathbf{C} \times \mathbf{D}_{\hbar}$ where $\mathbf{D}_{\hbar}$ is the cigar-like geometry of \cite{nw10} with the $\Omega$-deformation turned on. With an appropriate twist this geometry breaks half of the supersymmetries. Upon choosing the boundary conditions on $\mathbf{C} \times \p \mathbf{D}_{\hbar} = \mathbf{C} \times S^{1}$ which preserve the remaining supersymmetries, we can reduce our theory to a two-dimensional $\mathcal{N}=(2,2)$ theory on $\mathbf{C}$, the low energy description of which is characterized by the effective twisted superpotential
\beq
\mathcal{W}(\mathbf{a}, \mathbf{m}, \Lambda; \hbar) = \lim_{\ep \to 0} \ep \log{Z(\mathbf{a}, \mathbf{m}, \Lambda; \hbar)}
\eeq
For the perturbative contribution we have
\begin{multline}
\mathcal{W}^{pert.}(\mathbf{a}, \mathbf{m}, \Lambda; \hbar) = \lim_{\ep \to 0} \ep \log Z^{pert.}(\mathbf{a},\mathbf{m},\Lambda;\ep,\hbar)  =  \\ 
-\frac{1}{2 \hbar} \log \biggl( \frac{\Lambda^{2N}}{\hbar^{2N}} \biggr) \sum_{i=1}^{N} a_i^2 + \sum_{i,j=1}^{N} \varpi_{\hbar}(a_i - a_j) - \sum_{i=1}^{N}\sum_{a=1}^{N_f}  \varpi_{\hbar}(a_i - m_a)
\end{multline}
where $\varpi_{\hbar}(x)$ obeys
\beq
\label{varpi}
\frac{d}{dx} \varpi_{\hbar} (x) = \log \Gamma \biggl( 1 + \frac{x}{\hbar} \biggr) = const. - \sum_{n=1}^{\infty} \log \biggl(\frac{x+n \hbar}{\hbar} \biggr)
\eeq

The one-loop contribution $\varpi_{\hbar}(m)$ has a simple intuitive explanation as a contribution of infinite number of angular momentum modes with mass parameters $(m+n \hbar)$ for $n$-th mode into the effective twisted superpotential, which are chiral multiplets in the effective two dimensional theory. Indeed, after integrating out a single chiral multiplet we get \footnote{We use $\hbar$ for the characteristic scale of 2d theory.}
\beq
\Delta \mathcal{W}_n = -(m+n\hbar) \biggl[\log\biggl(\frac{m+n \hbar}{\hbar}\biggl)-1\biggr]
\eeq
and after summing over $n$ we get $\varpi_{\hbar}(m)$.

In principle, if no accidental cancellations happen, the effective two dimensional theory has infinitely many local degrees of freedom, which have different angular momentum along $\mathbf{D}$ weighted with the massive parameter $\hbar$. 

There are two natural boundary conditions \cite{nw10} which require
\beq
\Type A: \,\,\,\,\,\,\,\,\, \frac{a^{D}_i}{\hbar} = \frac{\p \mathcal{W}(\mathbf{a},\mathbf{m},\Lambda; \hbar)}{\p a_i} \in  \mathbf{Z}
\eeq
\beq
\Type B: \,\,\,\,\,\,\,\,  \frac{a_i}{\hbar} \in \mathbf{Z} + \frac{\theta_i}{2 \pi}, \,\,\,\,   \theta \in [0,2 \pi)
\eeq
Type $A$ corresponds to Neumann-type boundary condition for gauge fields leading to a dynamical vector multiplet in two dimensions. On the contrary type $B$ corresponds to Dirichlet boundary conditions, fixing the holonomy along the boundary of the cigar parametrized by $\theta_i$ and freezing gauge degrees of freedom.\footnote{There are could be more sophisticated boundary conditions which can be obtained e.g by coupling our four-dimensional gauge theory to some three dimensional theory on $\mathbf{C} \times  \p \mathbf{D}_{\hbar}$ in a supersymmetric fashion \cite{NPS13}.}  The choice of this boundary conditions specifies the quantization of an algebraic integrable system. 

The Bethe/gauge correspondence states that the vacua of the effective two dimensional theory are in one-to-one correspondence with the eigenstates of the Hamiltonians of the quantum integrable system. Moreover, the expectation values of the topologically protected chiral observables, which are traces of the adjoint scalars in a vector multiplet $\Tr \mathbf{\Phi}^k$ of a four dimensional theory, coincide with the eigenvalues of the Hamiltonians $\mathcal{H}_{k}$ on this states:
\beq
\langle vac | \mathcal{H}_{k} | vac \rangle
 \longleftrightarrow \langle \Tr \mathbf{\Phi}^k \rangle_{vac.}  
\eeq

The simplest example of such correspondence for purely four dimensional gauge theory is the periodic $A_{N-1}$ Toda chain.

\subsection{Pure $U(N)$ and the periodic Toda chain}
The (complexified) periodic Toda chain is a one-dimensional system of $N$ non-relativistic particles interacting with the following potential
$$
V(x_1, ... , x_N) = \Lambda^2 \sum_{i=1}^{N} e^{x_{i}-x_{i+1}}, \,\,\,\,\,\,\, x_{N+1} = x_{1}
$$
where the coordinates $x_i \in \mathbf{C}/(2 \pi \mathbf{Z})$ while the momenta $p_{i} \in \mathbf{C}$. The set of classical Hamiltonians $\{H_{i}\}$ is conveniently encoded in the spectral curve equation
\beq
\det(x-L(w)) = x^N + H_1 x^{N-1} + H_2 x^{N-2} + ... + H_N -\Lambda^N (z+z^{-1}) = 0
\eeq
where $L(z)$ is a Lax operator
\beq
L(z) = 
\left( \begin{matrix} 
p_1 & \Lambda^2 e^{x_1-x_2} & 0 & \dots & \dots & \Lambda^N  z^{-1} \\
1 & p_2 & \Lambda^2 e^{x_2-x_3} & \dots & \dots & 0 \\
0 & 1 & p_3 & \Lambda^2 e^{x_3-x_4} & \dots & 0 \\
0 & \dots & \dots & \dots & \dots & 0 \\
0 & \dots & \dots & \dots & p_{N-1} & \Lambda^2 e^{x_{N-1}-x_n} \\
\Lambda^{2-N} e^{x_N-x_1} z & 0 & \dots & 0 & 1 & p_N \\
\end{matrix} \right)
\eeq 
The first two Hamiltonians are 
\beq
H_1 = - \sum_{i=1}^{N} p_i
\eeq
\beq
H_2 = - \frac12 \sum_{i \neq j} p_i p_j + V(x_1, ... , x_{N})
\eeq
As is well known the underlying 4d $\mathcal{N}=2$ gauge theory for this system is the pure $U(N)$ theory. In particular its Seiberg-Witten curve coincides with the spectral curve of the integrable system \cite{gkmmm, warner}.

We are interested in the quantization of this system. The Hamiltonians and the momenta are now promoted to differential operators, acting on wave functions $\psi(x_1,..., x_N)$ with $p_i = \hbar \p_{i}$. There are two natural quantizations, corresponding to type $A$ and type $B$ boundary conditions \cite{NS09p3}. Type $A$ quantization corresponds to $x_i \in \mathbf{R}$ and $\psi(x_1- \bar{x}, ..., x_N - \bar{x}) \in L^2(\mathbf{R}^{N-1})$ where $\bar{x} = \sum_{i}^{N} x_i/N$ is the center of mass mode, which decouples in a trivial way. This condition leads to discrete unambiguous spectrum that corresponds to the set of vacua in the gauge theory, provided that the type $A$ boundary condition $a_{D}/\hbar \in \mathbf{Z}$ is satisfied. 

In this paper we are interested in type $B$ quantization that corresponds to $x_i \in i \mathbf{R}/2 \pi \mathbf{Z}$ and quasi-periodic non-singular wave functions 
\beq
\psi(x_1, ... , x_a + 2 \pi i, ... , x_N) = e^{i \theta_a} \psi (x_1, ..., x_N)
\eeq
The quasi-periodicity parameters $\theta_{a}\in [0,2\pi)$ are also known as Bloch-phases. In the special case of $N=2$, after the decoupling of the center of mass mode, the equation on the wave function coincided with Mathieu equation
\beq
-\hbar^2 \psi''(x) + 8 \Lambda^2 \cos(2 x) \psi(x) = 8 u \psi(x)
\eeq
where $u = \frac14 \langle \Tr \mathbf{\Phi}^2 \rangle$. For real $\Lambda$ and $\hbar$ it describes a particle moving in a periodic cosine potential. At fixed $\theta$-parameters the spectrum is discrete. However as we vary them the spectrum consists of peculiar structure of bands and gaps. In particular for small $\Lambda$ and when we sit near the edge of some band, the spectrum has exponentially small in $\sim 1/\hbar$  splitting of the eigenvalues. More precisely if $\theta_1 = ... = \theta_{k_1}; \theta_{k_1+1} = ... = \theta_{k_2}; ... ; \theta_{k_m+1} = ... = \theta_{k_N}$ and are all integers, then instead of naive $\frac{N!}{k_1!k_2!...k_N!}$-fold degeneracy as for $\Lambda=0$ we have non-degenerate spectrum due to quantum reflection on the potential.

This second type of quantization appears to be more mysterious from the gauge theory point of view. When all $\theta$-parameters are zero $a_a/\hbar$ are forced to be equal to integer, and when $a_{ab}/\hbar \in \mathbf{Z} \setminus \{0\}$ some perturbative $W$-boson modes naively become massless that is clear from the logarithmic singularities in their perturbative contribution into the effective twisted superpotential. Thus naively the effective description has to break down at this locus, as was pointed out in \cite{Jeong17}. One of the main purposes of the present paper is to resolve this puzzling issue.

\section{Effective two-dimensional field theory}
In this section we are interested in the physics of the effective two-dimensional $\mathcal{N}=(2,2)$ gauge theory, which appears upon the reduction of $\mathcal{N}=2$ four-dimensional gauge theory in the $\Omega$-background as was described in the previous section. Our main example will be the simplest non-trivial case of pure $U(2)$ gauge theory, though we will also comment on a few generalizations.

\subsection{Pure $U(2)$ theory}
In this case the superpotential $\mathcal{W}(a, \Lambda; \hbar)$ depends on a single complex parameter $a$ parametrizing vacua.
The superpotential has the following expansion:
\beq
\mathcal{W}(a, \Lambda; \hbar) = \mathcal{W}^{pert.}(a, \Lambda; \hbar) - \hbar \, F(\nu, q)
\label{superpot}
\eeq
\beq
F(\nu,q)= \sum_{k=1}^{\infty} F_{k}(\nu) q^{2k}
\eeq
where we have introduced dimensionless variables $\nu$ and $q$: 
\beq
\nu = {2a \over \hbar}, \,\,\,\,\,\,\,\,\, q = \frac{\Lambda^2}{\hbar^2}
\eeq	

The perturbative part $\mathcal{W}^{pert.}$ of the superpotential is
\beq
\label{Wpert}
\mathcal{W}^{pert.} (a, \Lambda; \hbar) = - \frac{2 a^2}{\hbar} \log \biggl( \frac{\Lambda^2}{\hbar^2} \biggr)  + \varpi_{\hbar}(2a) + \varpi_{\hbar}(-2a)
\eeq
and the first few terms for $F_{k}(s)$ are 
\begin{multline}
\sum_{k=1}^{\infty} F_{k}(\nu) q^{2k} = \frac{2}{\nu^2-1} q^2 + \frac{5 \nu^2 +7}{(\nu^2-1)^3(\nu^2-4)} q^4 + \frac{16 \left(9 \nu^4+58 \nu^2+29\right)}{3 \left(\nu^2-1\right)^5 \left(\nu^2-4\right)(\nu^2 - 9)} q^6 \\+ \frac{1469 \nu^{10}+9144 \nu^8-140354 \nu^6+64228 \nu^4+827565 \nu^2+274748}{2 \left(\nu^2-4\right)^3 \left(\nu^2-1\right)^7 \left(\nu^2-9\right)  \left(\nu^2-16\right)} q^8 + ...
\label{eq:oldF}
\end{multline}

The characteristic feature that we observe for all $F_{k}(s)$ is that they all have poles at non-zero integers $s=n$ for $-k \leq n\leq k$. However, as we will see shortly they are just artifacts of the expansion in small $q$. 

There is a single independent chiral trace operator $u = \frac{1}{4} \langle \Tr \mathbf{\Phi}^2 \rangle$. It is related to the superpotential via quantum version of Matone relation \cite{Matone95}, \cite{FFMP04}, \cite{Whitham}
\beq
u = - \frac{\hbar}{8} \Lambda \frac{\p \mathcal{W}(a,\Lambda; \hbar)}{\p \Lambda} 
\label{eq:TMatone}
\eeq

Below we will often need the monopole mass $|a_D^M|$ which is in our normalization\footnote{One can fix the normalization by requiring that $a$ and $a_D^M$ are given by 
the integral of the Seiberg-Witten form $pdq = \frac{\sqrt{2}}{2\pi} \sqrt{u-\Lambda^2 \cos{q}}\ dq$ over the corresponding A-cycle $[-\pi,\pi]$ and B-cycle 
$[\arccos{u/\Lambda^2},2\pi - \arccos{u/\Lambda^2}]$}  reads as:
\beq
\label{adm_def}
a_D^M = i\cfrac{\hbar}{4 \pi} \cfrac{\pr \mathcal{W}(a,\La;\hbar)}{\pr a}
\eeq
For large $a$ such that $a \gg \La$  and $a \gg \hbar$ instanton corrections are suppressed and one can find using eqs. (\ref{Wpert}) and (\ref{varpi}) that:
\beq
\label{adm}
a_D^M = i\cfrac{2a}{\pi} \log{\cfrac{2a}{\Lambda}} + O(a)
\eeq
since in this limit $\varpi_{\hbar}(x) = \frac{x^2}{2 \hbar} \log{\frac{x}{\hbar}} + O(x) $
\subsubsection{Mathieu equation}
As was discussed in the previous section, the finding of the expectation values of chiral trace operator in different vacua of the effective two dimensional theory for type $B$ boundary conditions is equivalent to finding of the spectrum of the Mathieu equation
\beq
-\psi''(x) + 8 q \cos(2 x) \psi(x) = \frac{8 u}{\hbar^2} \psi(x)
\eeq
with quasi-periodic boundary conditions $\psi(x+ \pi) = e^{\pi i \nu}\psi(x)$. In particular, for real $\Lambda$ and $\hbar$  the spectrum has an alternate set of 
bands and gaps with exponentially small band widths in ``low'' spectrum when $u \approx -\Lambda^2$ and 
exponentially small gap widths in ``high'' spectrum when $u \gg \Lambda^2$. 

\begin{figure}[h!]
\centering
\includegraphics{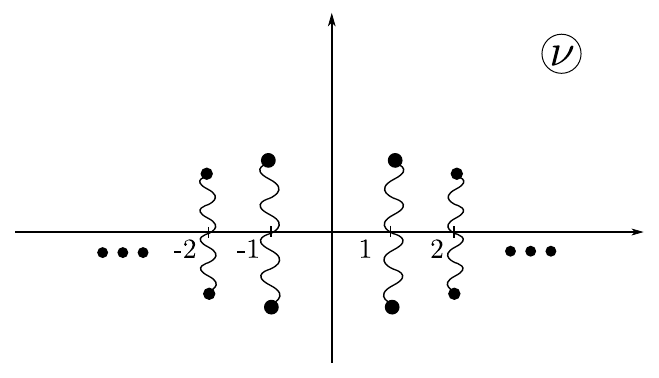}
\caption{This picture shows the branch cut structure of the function $u(q,\nu)$. Each cut near $\nu = n$ is identified with the cut near $\nu=-n$. As the result of this identification we obtain an infinite-genus surface.}
\label{fig:plane}
\end{figure}

Since $\pi \nu$ is a Bloch phase the function $u(q,\nu)$ has period $2$ in $\nu$ on the real axes. 
However, this periodicity is not manifest in the expansion obtained by inserting (\ref{eq:oldF}) 
into the Matone relation (\ref{eq:TMatone}). For example, when we change $\nu$ along $[0,2]$ and 
are supposed to return back we meet the singularity at $\nu=1$ in each term of the expansion. 
A possible resolution is that instead of these naive poles the function $u(\nu,q)$ has branch cuts as shown on the figure (\ref{fig:plane}) with the identification of cuts near $\nu = n$ and $\nu = -n$ to make it 2 periodic, and when we expand it in $q$ we loose this branch cut structure and obtain poles. The latter happens if the widths of this cuts vanish as $q$ goes to 0. The identification of $\nu = n$ and $\nu = -n$ is natural since $\nu$ is proportional to a Coulomb parameter $a$ which transforms under Weyl symmetry as $a \to -a$. 

\subsubsection{Resummation}

To obtain a good expansion of $F(\nu,q)$ near the cuts we use the same trick as in \cite{Becc} and reorganize the expansion in $q$ in the following way
\beq
F(\nu, q) = \sum_{n>0} \sum_{k=1}^{\infty} \biggl[g_{k}^{(n)}\biggl( \frac{q^{n}}{n-\nu} \biggr) +g_{k}^{(n)}\biggl( \frac{q^{n}}{n+\nu}\biggr) \biggr] q^{2k - 2 + n} 
\label{eq:newF}
\eeq
where $g_k^{(n)}(z) = \mathcal{O}(z)$ and we have introduced variable $z$: 
\beq
z = \frac{q^n}{n-\nu}
\eeq
What is physical meaning of this resummation?  From more technical point of view, it corresponds to resumming separately the leading singularity from each
instanton sector(function $g_1^{(n)}$) then the subleading singularity (function $g_2^{(n)}$) and so on. 
As we will see shortly, $\log{q^n}$ plays the role of 2d FI parameter, whereas $n-\nu$ is proportional to the mass of light W-boson mode.
The combination $\log{q^n}-\log \l n-\nu \r$ is exactly 2d effective FI parameter once we integrate out a particle of mass $n-\nu$. Therefore this expansion 
reorganizes the 4d instanton expansion into a double expansion in terms of 2d effective vortices and 4d instantons.

By comparing the first few terms in the expansions of (\ref{eq:newF}) and $(\ref{eq:oldF})$ in $q$ we expect that $g_k^{(n)}$ have the following form
\beq
\label{eq:gn}
g^{(n)}_{1} (z) = \frac{\log \left(\frac12+\sqrt{\frac14+\frac{z^2}{\zeta_n^2}} \right)+1-\sqrt{1+\frac{4 z^2}{\zeta_n^2}}}{z} 
\eeq
for $k=1$ and

\beq
g_{k}^{(n)}(z) = \frac{\left(1+\frac{4 z^2}{\zeta_n^2}\right)^{\frac{5}{2}-k} Q_{2 k-3}\left(z^2\right)  + P_{k-1}\left(z^2\right)}{z^{2 k-1}}
\eeq
for $k>1$, where $\zeta_{n} = n! (n-1)!$ and $P_m(w)$, $Q_m(w)$ are some polynomials of degree $m$, specific for each $n$ and $k$. Using this ansatz and $g_k^{(n)}(z) = \mathcal{O}(z)$, one can compute $P$ and $Q$ polynomials which appear $g_k^{(n)}(z)$ for any given $n$ and $k$ by expanding the Nekrasov partition function up to the corresponding order. We present the first few $g_k^{(n)}$ functions in the Appendix \ref{sec:gs}.

Once we have reorganized the expansion in this form, we capture the branch cut structure provided that we identify the cuts near $n$ and $-n$ that was proposed in the previous subsection. In particular, we can write down a perturbative in $q$ expressions for band/gap edges for $\nu \approx n$ using the quantum Matone relation, which coincide with the known expressions (see e.g. \cite{DB15})

\beq
\begin{split}
u_0 = {\hbar^2 \over 8} \l - q + \frac{7}{4} q^3 -\frac{58}{9} q^5 + \frac{68687}{2304} q^7+\ldots \r \\
u_1^- = {\hbar^2 \over 8} \l 1  - 4 q - 2 q^2 + q^3 - \frac{1}{6} q^4 - \frac{11}{36} q^5 +\ldots \r \\
u_1^+ = {\hbar^2 \over 8} \l 1  + 4 q - 2 q^2 - q^3 - \frac{1}{6} q^4 + \frac{11}{36} q^5+\ldots \r \\
\dots
\end{split}
\label{eq:energies}
\eeq

\subsubsection{Local model}

Now let us try to explain the physical implications of the appearance of the branch cuts instead of the naive poles. Let us  build a simplified local model near the first pole $m=(\hbar - 2 a) \to 0$ in the limit of very large $\Omega$-background $\hbar \gg \Lambda$ or $q \ll 1$. Naively, when $\hbar$ and $a$ are very large, the instanton corrections are switched off and only the perturbative part contributes
\beq
\mathcal{W}^{pert.}(a,\Lambda; \hbar) \approx  -m \biggl[\log\biggl(\frac{m}{\hbar} \biggl)-1\biggr] + const.
\eeq
which is nothing else but the contribution of a first W-boson angular momentum mode, that naively becomes massless at $2a = \hbar$. However, looking at the reorganized expansion for the non-perturbative part we see that the first leading term has the same order in $q$ as the perturbative one, and 

\begin{multline}
\mathcal{W}(a, \Lambda; \hbar) \approx   - m \biggl[\log\biggl(\frac{m}{\hbar} \biggl)-1\biggr] - q  \hbar \,  g^{(1)}_{1}  \biggl( \frac{q}{1-\nu}\biggr) + const. \\
=  - m \log \biggl(\frac{m+\sqrt{m^2+\frac{4 \Lambda^4}{\hbar^2}}}{2 \hbar} \biggr) +\sqrt{m^2+\frac{4 \Lambda^4}{\hbar^2}} + const.
\label{eq:locW}
\end{multline}

It is instructive to look at the mirror description of this system. Consider 
\beq
\mathcal{W} = m (Y - t_{eff.}) + e^{-Y} + (\Lambda^4/\hbar^2)e^{Y}.
\label{eq:locMirr}
\eeq 
where $t_{eff.} = \log (\hbar^{-1})$. If we integrate out the field $Y$ we return back to (\ref{eq:locW}). What we obtain is the mirror description of chiral multiplet deformed by $(\Lambda^4/\hbar^2) e^{Y}$ operator. 

Now we can try to give a physical interpretation of this additional term. As is well known $e^{-Y}$ is related to the appearance of vortex configurations in the mirror description. Similar $e^{Y}$ is related to the appearance of vortex configurations with the opposite charge. In the case of free chiral multiplet such term doesn't appear since it has wrong quantum numbers. However in our case the parameter $(\Lambda^4/\hbar^2)$ also carries quantum numbers and in combination with $e^{Y}$ allows for the terms of the form $(\Lambda^4/\hbar^2)^{n} e^{(2n-1) Y}$. Since we work in the first non-zero order in $(\Lambda^4/\hbar^2)$ we have only the term $(\Lambda^4/\hbar^2) e^{Y}$. The appearance of $\Lambda^4$ which measures the instanton charge suggests that the configuration that has a negative vortex charge also carries the instanton charge and it is this charge that allows for negative vortex numbers.

Curiously similar deformation appears in a local model of a theory of a chiral doublet surface defect in a pure $SU(2)$ gauge theory, which is obtained by coupling 2d $\mathcal{N}=(2,2)$ chiral doublet theory, living on $\mathbf{C} \subset \mathbf{C}^2$, to 4d $\mathcal{N}=2$ $SU(2)$ theory via weakly gauging of a global $SU(2)$ symmetry of the doublet \cite{GGS13}.
In that case the effective twisted superpotential for the defect, that takes into account bulk effects, is
\beq
\mathcal{W} = - \langle \Tr (m+\Phi) \bigl( \log (m+ \Phi) - 1\bigr) \rangle_{4d}
\eeq
where $m$ is the twisted mass parameter for the diagonal $U(1)$ symmetry of the doublet and $\Phi$ is the bulk vector multiplet scalar. The bulk averaging $\langle ... \rangle_{4d} $ can be done using the resolvent of \cite{NO02}
\beq
R(x) = \Tr \frac{1}{x + \Phi} = \frac{2 x}{\sqrt{(x^2-2u)^2 - 4 \Lambda^2}}
\eeq
which gives 
\beq
- \p_m \mathcal{W} = \log{\Lambda^2} + \arccosh \frac{m^2 - 2u}{2 \Lambda^2}
\eeq
and for large $m^2$ and $2u \approx a^2$ close to $2u \approx m^2$
\beq
- \p_m \mathcal{W} \approx \log{\Lambda^2} + \arccosh \frac{m - a}{\Lambda^2/a}.
\eeq
A mirror description of this region is 
\beq
\mathcal{W} = (m-a) Y + e^{-Y} - (\Lambda^4 / 4 a^2) e^{Y} 
\eeq
and similarly to our case the perturbative 2d description of the chiral doublet is deformed by $(\Lambda^4/(4a^2)) e^{Y}$ which appears due to instantons in the bulk. In this situation the source of this term has a simple interpretation in the brane construction of this defect \cite{Gai09} and appears in the microscopic derivation of the superpotential \cite{GLMS17, lerda1, lerda2}.

\subsubsection{Wall-crossing near the cut}
Now we can answer what happens with the $W$-boson mode that naively becomes massless near $\nu \approx n$. It has to disappear from the BPS spectrum and thus decays into other BPS objects on some curve of marginal stability.

Let us again consider the mirror superpotential for our local model (\ref{eq:locMirr}) near the first pole. The vacuum values of $Y$ are
\beq
Y^{(\pm, n)} = \log \biggl( \frac{-m \pm \sqrt{m^2+\frac{4 \Lambda^4}{\hbar^2}}}{\frac{2 \Lambda^4}{\hbar^2}}\biggr)+ 2 \pi i n , \,\,\,\,\, n \in \mathbb{Z}
\eeq
where $2 \pi i n$ appears due to the multivaluedness of the logarithm function. In the original global description this vacua correspond to two split states near $\nu \approx 1$ and lie in the two neighboring bands. It is easy to find the degeneracies of BPS particles, connecting different vacua using the  methods of \cite{CV92}, \cite{GMN11}. However one can simply notice that our local superpotential coincides with the well known mirror description of $\mathbf{CP}^1$ sigma model with a twisted mass, for which the BPS spectrum was already analyzed in \cite{dorey98}. Indeed, the twisted superpotential for $\mathbf{CP}^1$ has the form (e.g. \cite{HV00})
\beq
\mathcal{W}_{\mathbf{CP}^1}(t,m) = \Sigma (Y_1 + Y_2 - 2 t) + \frac{m}{2} (Y_1-Y_2) + e^{-Y_1} + e^{-Y_2}
\eeq
where $t$ and $m$ are the K\"ahler parameter of $\mathbf{CP}^1$ and the twisted mass, correspondingly. After integrating out the fields $\Sigma$ and $Y_2$ we obtain
\beq
\mathcal{W}_{\mathbf{CP}^1}(t,m) =  m (Y_1- t) + e^{-Y_1} + e^{-2 t} e^{Y_1}
\eeq
that gives us (\ref{eq:locMirr}) up to a constant if $e^{-t} = \Lambda^2/\hbar$.

Translating \cite{dorey98} into our notations, the result for the BPS spectrum is the following. In the region of large enough $m$ there is a single state that connects $Y^{(\pm,n)}$ and $Y^{(\pm,n+1)}$ and a tower of single states which interpolate between $Y^{(+,n)}$ and $Y^{(-,n+k)}$. The former one has a mass $|m|$ and corresponds to a W-boson mode while the latter corresponds to solitons, coupled to $k$ W-boson modes. On the other hand in the region of small $m$ there are only two single states, connecting vacua $Y^{(-,n)}$ and $Y^{{(+,n)}}$. The masses of this two particles are
\beq
M_{sol.} = \biggl| \frac{1}{2\pi} \frac{\Lambda^2}{\hbar} \biggl[ 4 \sqrt{1+\frac{\hbar^2 m^2}{4 \Lambda^4}} + \frac{\hbar m}{\Lambda^2} \log \biggl( \frac{m \hbar - \sqrt{4 \Lambda^2 + m^2 \hbar^2}}{m \hbar + \sqrt{4 \Lambda^2 + m^2 \hbar^2}} \biggr) \biggl] \biggr|.
\eeq
and in the $m \ra 0$ limit
\beq
M_{sol.} = \biggl| \frac{2}{\pi} \frac{\La^2}{\hbar} \biggr|
\eeq
On the curve of marginal stability $|m| = 2 M_{sol.}$ all states of the large $m$ region including W-boson mode decay into this two particles.

Note that the electric charge of W-boson in the 4d theory is 2 and the electric charge of all other BPS particles is integer in the units of W-boson charge. On the other hand in the effective 2d theory the solitons both have charge 1. Thus this solitons do not simply come from the modes of other BPS particles but appear only due to the generated superpotential.   

{\bf Remark.} One may ask why we do not obtain in the spectrum of 2d BPS particles the modes of 4d particles with non-zero magnetic charge. At least in the undeformed theory in the region of large $u$ we expect the whole tower of dyons, but the effective twisted superpotential doesn't have monodromies associated with this particles. The resolution of this puzzle lies in the boundary condition that we impose. Indeed as was shown in \cite{nw10} type $B$ quantization leads to Dirichlet boundary condition for gauge fields. As the result the flux of magnetic field through $\mathbf{S}^1_{\infty} \times \p \mathbf{D}_{\hbar}$ where $\mathbf{S}^1_{\infty}$ is the circle at infinity of $\mathbf{C}$ has to be zero and that select only the particles with zero magnetic charge. In contrast the same reasoning allows only for electrically neutral particles for type $A$ quantization. 

Though this analysis was done for the first gap $2a \to \hbar$, it is straightforward to do the same local analysis near all other gaps for small $q$. 
We again obtain a pair of solitons with masses 
\beq
 M_{n} = \frac{2 \hbar}{\pi}  \frac{1}{n!(n-1)!} \biggl(\frac{\Lambda^2}{\hbar^2} \biggr)^n
\eeq
and $n$-th $W$-boson mode decays into this solitons near the corresponding gap. Note that for large $n$ using eq. (\ref{adm}) we have:
\beq
\label{mnsol}
M_n \sim \exp \biggl( - \frac{2 \pi |a^M_D|}{\hbar} \biggr)
\eeq

However, our analysis is restricted to the limit of large $\Omega$-background $\La/\hbar \ll 1$  and to a set of light BPS particles. In particular it doesn't tell us 
anything about the strong coupling region $u \sim \Lambda^2$. 
It would be interesting to explore this region and the global structure of the BPS spectrum.

\subsection{Pure $U(N)$ theory}

So far we have considered the case of $U(2)$ gauge theory and have seen, that $\mathcal{N}=(2,2)$ $\mathbf{CP}^1$ model appears in the local description "near the gap". One may ask how it generalizes to the case of $U(N)$ theories, e.g. for $N$-particle periodic Toda system. In that case we already have different types of singularities parametrized by a set of integers $(k_1, ..., k_l)$ such that $k_1+...+k_l = N$, as was described in section 2. This singularities appear at $a_1 =...=a_{k_1}$, $a_{k_1+1}=...=a_{k_2}$, ... , $a_{k_{l-1}+1}=...=a_{k_{l}}$ and $(a_{a}-a_{b}) \in \mathbf{Z} \setminus \{0\}$ otherwise. The degeneracy of this singularities is $\frac{N!}{k_1! ... k_l!}$. 

A natural candidate for the local description near the singularity is a $\mathcal{N}=(2,2)$ sigma model on the space $\Fl_{(k_1,...,k_l)}$ of flags $\mathbf{C}^{k_1} \subset \mathbf{C}^{k_1+k_2} \subset ... \subset \mathbf{C}^{N}$. As a trivial check the degeneracy of the singularity coincides with the number of vacua in this sigma model, and the number possible deformations of Coulomb parameters $\mathbf{a}$ coincides with the number of possible twisted mass parameters.   However, in this sigma model we can also vary $l-1$ K\"ahler parameters $\{t_s\}$, the option that we don't have in the local model. So the K\"ahler parameters have to be specified in terms of $\Lambda$ and $\hbar$.

\begin{figure}
\centering
\begin{tikzpicture}[thick,
  ->-/.style 2 args={decoration={markings, mark=at position #1 with
{\arrow[rotate=#2]{stealth}}},postaction={decorate}}]
  \node[circle,draw,minimum size=2.5em] (1) at (0,0) {$n_1$};
  \node[circle,draw,minimum size=2.5em] (2) at (2,0) {$n_2$};
  \node (dots) at (3.5,0) {$\cdots$};
  \node[circle,draw,minimum size=2.5em] (M1) at (5,0) {$\mathclap{n_{l-1}}$};
  \node[draw,minimum size=2.5em] (N) at (7,0) {$N$};
  \draw[->-={.65}{0}] (1) -- (2);
  \draw[->-={.65}{0}] (2) -- (dots);
  \draw[->-={.65}{0}] (dots) -- (M1);
  \draw[->-={.65}{0}] (M1) -- (N);
\end{tikzpicture}
\caption{A quiver diagram for the gauged linear sigma model that is described by the sigma model on a flag variety.}
\label{fig:defect_quiver}
\end{figure}
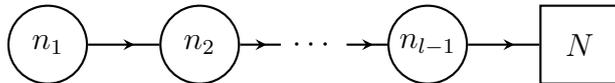

The effective twisted superpotential for this sigma model can be obtained via describing it as a low-energy limit of a gauged linear sigma model. The set of fields is encoded in a quiver shown on the figure (\ref{fig:defect_quiver}). 

Each circle-node corresponds to a vector multiplet for a gauge group $U(n_s)$ where $n_s = k_1+...k_s$ and each arrow corresponds to a chiral multiplet in the bifundamental representation. The last square-node corresponds to a global $U(N)$ symmetry with parameters $m_1, ... , m_N$, which correspond to the twisted mass parameters of the sigma model. The K\"ahler parameters comes from the Fayet-Iliopoulos parameters $t_s$ on each node. The effective superpotential has the following form \footnote{Here we omit an obvious characteristic scale of 2d theory.}
\begin{multline}
\mathcal{W}(\mathbf{\sigma}, \mathbf{m}, \mathbf{t}) = \sum_{s=1}^{l-1} \sum_{i=1}^{n_s} t_s \sigma^{(s)}_i - \sum_{i=1}^{N}\sum_{j=1}^{n_{l-1}} (m_i - \sigma^{(l-1)}_j) \log \frac{(m_i - \sigma^{(l-1)}_j)}{e} \\ 
- \sum_{s=1}^{l-2} \sum_{i=1}^{n_s}\sum_{j=1}^{n_{s+1}} (\sigma^{(s+1)}_i - \sigma^{(s)}_j) \log \frac{(\sigma^{(s+1)}_i - \sigma^{(s)}_j)}{e} \\ 
+ \sum_{s=1}^{l-1} \sum_{i \neq j}^{n_s} (\sigma^{(s)}_i - \sigma^{(s)}_j) \log \frac{(\sigma^{(s)}_i - \sigma^{(s)}_j)}{e} \\ 
\label{eq:SMpot}
\end{multline}
where $\sigma^{(s)}_i$ are the Coulomb parameters in the vector multiplet of the $s$-th node. After minimizing it with respect to $\mathbf{\sigma}$ by solving vacuum equations
\beq
\exp \frac{\p \mathcal{W}(\mathbf{\sigma}, \mathbf{m}, \mathbf{t})}{\p \sigma_i} \biggl|_{\sigma_*} = 1
\eeq
 we obtain the effective twisted superpotential for our sigma model $\mathcal{W}_{SM} (\mathbf{m},\mathbf{t})= \mathcal{W}(\sigma_*, \mathbf{m},\mathbf{t})$.

We expect that after proper identification of $\mathbf{t}$ in terms of $\Lambda$ and $\hbar$ the result of the resummation of the instanton partition function near the corresponding singularity $\tilde{\mathbf{a}}$ will be given by $\mathcal{W}_{SM}(\mathbf{m},\mathbf{t})$ with $m_i = \tilde{a}_i - a_i$ . In the next subsection we make make a non-trivial check of this proposal for the case of the full flag variety  $\Fl_{(1,....,1)}$ near the singularity $\tilde{a}_i = (i-1)\hbar$.  

Our strategy is the following. We turn on the second $\Omega$-background parameter $\ep$ and consider the fully deformed Nekrasov partition function. From the point of view of the effective two dimensional theory it corresponds to introducing a two-dimensional $\Omega$-background. The resulting partition function is the vortex partition function \cite{Shadchin06, NekrQuarks, DGH10, DGLL12} for this effective theory. We will see that in the limit $\hbar \to \infty$ near the gap this partition function reduces to the vortex partition function of the desired sigma model.

\subsubsection{Full flag sigma model from Nekrasov partition function}

Let us consider the Nekrasov partition function of pure $U(N)$ gauge theory with $a_s = (s-1) \hbar - m_s$ in the limit $\hbar \to \infty$ with $z = \Lambda^{2N}/\hbar^{2N-2}$ fixed. Looking at the character of the tangent space at a fixed point $\lambda$
\beq
\mathcal{T}^{pure}_{\lambda} = \mathcal{W} \mathcal{V}_{\lambda}^* + e^{\ep+\hbar} \mathcal{V}_{\lambda} \mathcal{W}^* - (1 - e^{\ep}) (1-e^{\hbar}) \mathcal{V}_{\lambda} \mathcal{V}_{\lambda}^*
\eeq
one can see that only a subset of diagrams with the condition that $\lambda_s$ has at most $(N-s)$ rows has a 
non-vanishing contribution. Indeed, if this condition is not satisfied then too many terms terms in the character contain $\hbar$ in the exponent that leads to a $o(\hbar^{2-2N})$ contribution after applying $e$-symbol.

Let us introduce $k^{(s)}_i = \lambda_{i,s-i+1}$, $\sigma^{(s)}_i = m_i - \ep k^{(s)}_i$ and 
$$z_s = \frac{(-1)^N z}{s!(s-1)!(N-s)!(N-s-1)!}$$
Then the Nekrasov partition function takes the following form 
\begin{multline}
Z^{inst.} = \sum_{k^{(s)}_i \geq 0} \biggl( \prod_{s=1}^{N-1} z_s^{\sum_i k^{(s)}_i} \times\prod_{s=1}^{N-1} \prod_{i=1}^{s} \prod_{j=1}^{s+1} \frac{1}{(\sigma^{(s+1)}_j - \sigma^{(s)}_i; - \ep)_{k^{(s)}_i-k^{(s+1)}_j} } \\
\times \prod_{s=1}^{N-1} \prod_{i \neq j}^{N} (\sigma^{(s)}_i-\sigma^{(s)}_{j}; - \ep)_{k^{(s)}_j-k^{(s)}_i} \biggr)
\end{multline}
where 
$$(a;\ep)_n = \ep^n \frac{\Gamma(n+a/\ep)}{\Gamma(a/\ep)} = a(a+\ep)...(a+(n-1)\ep)$$
and $\sigma^{(N)}_i = m_i$. This is precisely the vortex partition function for $\Fl_{(1,1,...)}$ sigma model with exponentiated FI parameters $z_i = e^{-t_i}$. The first product comes from the contribution of bifundamental chiral multiplets and the fundamental chiral multiplet on the last node. The second product is the contribution of the vector multiplets.

In the limit $\ep \to 0$ 
\beq
Z^{inst.} \sim e^{\frac{1}{\ep} \mathcal{W}(\sigma, \mathbf{m}, \mathbf{t})} 
\eeq
and using 
\beq
(a;\ep)_{(\sigma-a)/\ep} \sim e^{-\frac{1}{\ep} \bigl(a \log \frac{a}{\ep} - a \bigr)  + \frac{1}{\ep}\bigl(\sigma \log \frac{\sigma}{\ep} - \sigma \bigr)}
\eeq
we obtain (\ref{eq:SMpot}) from the vortex partition function up to the perturbative contribution. The latter comes from the perturbative superpotential $\mathcal{W}^{pert.}(\mathbf{a},\Lambda; \hbar)$.

\subsection{$\mathcal{N}=2^*$ theory}
One natural deformation of our story can be obtained by considering $U(N)$ four dimensional theory with the matter multiplet in the adjoint representation, that is known as $\mathcal{N}=2^*$ theory. It's mass parameter $m$ plays the role of the deformation parameter.

Its instanton partition function is 
\beq
Z^{inst.} (\mathbf{a} ,m, \Lambda; \ep, \hbar) = \sum_{\{\lambda\}} q_{\tau}^{|\lambda|} e( \mathcal{T}^{\mathcal{N}=2^*}_{\lambda})
\eeq
where
\beq
\mathcal{T}^{\mathcal{N}=2^*}_{\lambda} = (1-e^{m})\mathcal{T}^{pure}_{\lambda}
\eeq
and the instanton counting parameter is $q_{\tau} = e^{2 \pi i \tau}$. In the decoupling limit $m \to \infty$ with a proper rescaling of the instanton counting parameter $\Lambda^{2N} = q_{\tau} m^{2N}$ we return back to a pure gauge theory.

The underlying integrable system is also well-known. It is the elliptic Calogero-Moser model which is the system of particles on a circle with the Weierstrass $\wp$-function potential with the modular parameter $\tau$. Its second quantum Hamiltonian has the form  
\beq
\hat{H}_2 = -\frac{\hbar^2}{2} \sum_{i=1}^{N} \p_i^2 + m (m + \hbar) \sum_{i<j} \wp(x_i-x_j)
\eeq
and for $N=2$ the spectral problem coincides with the spectral problem of Lame equation
\beq
-\frac{\hbar^2}{2} \psi''(x) + m(m + \hbar) \wp(x) \psi(x) = - u \psi(x)
\eeq

The quantized problem clearly resembles the case of Toda systems with the same issue of the appearance of gaps and bands in the spectrum. The same methods of resummation can be applied here providing the effective description in the weak coupling limit at large $\hbar$. 

The superpotential is
\beq
\mathcal{W}(a, m, q_{\tau}; \hbar) = \mathcal{W}^{pert.}(a, m, q_{\tau}; \hbar) - \hbar \, F(\nu, \mu, q)
\label{superpot2}
\eeq
where $\mu = m/\hbar$, $q = q_{\tau}^{1/2}$ and 
\begin{multline}
\mathcal{W}^{pert.}(a,m,q_{\tau};\hbar) = - \frac{2 a^2}{\hbar} \log q  + \varpi_{\hbar}(2a) + \varpi_{\hbar}(-2a) \\
-  \varpi_{\hbar}(m+2a) - \varpi_{\hbar}(m-2a)
\end{multline}
The non-perturbative part after the reorganization has the following form
\beq
F(\nu,\mu, q) = F_0(\mu,q) + \sum_{n>0} \sum_{k=1}^{\infty} \biggl[g_{k}^{(n)}\biggl( \frac{q^{n}}{n-\nu},\mu \biggr) +g_{k}^{(n)}\biggl( \frac{q^{n}}{n+\nu},\mu\biggr) \biggr] q^{2k - 2 + n} 
\label{eq:newF2}
\eeq
By computing the first few $g^{(n)}_1$ functions one can show that
\beq
g^{(n)}_{1} (z) = \frac{\log \left(\frac12+\sqrt{\frac14+\frac{c_n^2 z^2}{\zeta_n^2}} \right)+1-\sqrt{1+\frac{4 c_n^2 z^2}{\zeta_n^2}}}{z} 
\eeq
where
\beq
c_n = \prod_{k=-n+1}^{n}(\mu+n)
\eeq
For generic $\mu$ the effective description near $\nu \approx n$ is the same as for pure $U(2)$ gauge theory up to renormalization of $q^n \to c_n q^n $. 

A new feature appears for $\mu \in \mathbf{Z}_{\geq0}$ (or $\mu \in \mathbf{Z}_{< 0}$) when $c_n = 0$ for $n > \mu$. In that case the corresponding $g$-functions become trivial. However, precisely at this point the perturbative contribution of matter multiplet cancels the logarithmic singularity of the W-boson mode, leading to a regular behavior near this points. This is in agreement with the well known fact that for such special $\mu$ the Lame equation has finite amount of gaps.

\section{Non-perturbative gaps in the QM spectrum and classical Liouville theory}

\subsection{Twisted superpotentials versus classical conformal blocks}

In this Section we shall use the AGT \cite{agt} and $AdS_3/CFT_2$ holographic correspondence to identify 
our findings regarding the mechanism of non-perturbative gap formation in QM. We will address 
the classical limit of the Liouville theory and a
scattering process in the $AdS_3$ gravity involving heavy  operators. Two groups of questions can be formulated

\begin{itemize}

\item What is the meaning of the exponentially small gaps in the 
QM spectrum from the classical  conformal block viewpoint? What
kind of non-perturbative phenomena in $AdS_3$ gravity  this exponentially small
factor captures if any?

\item What is the meaning of the CMS near the naive poles in the 
classical conformal block in the Liouville theory? This CMS can be
equally considered as the curve on the plane
of intermediate conformal dimension at fixed time or in the complex
time plane at fixed intermediate dimension. What is the meaning 
of the CMS curve from the viewpoint of $AdS_3$ scattering process
involving some non-perturbative contribution?

\end{itemize}
We shall present the qualitative discussion of these questions in this Section
postponing the detailed analysis for the separate publication.

Let us briefly remind that according to the AGT correspondence the Nekrasov partition function
is identified with the particular conformal block in the Liouville theory whose
type depends on the matter content on the gauge theory side. The coordinate on the
Coulomb branch corresponds to  dimension of the intermediate state in the 
conformal block therefore the poles we are hunting for correspond to the particular
values of the intermediate dimensions. 
The central charge in the Liouville theory
is expressed in terms of the parameters of the $\Omega$-deformation 
as follows
\beq
c= 1+6Q^2,\qquad Q=b+ \frac{1}{b},\qquad b^2=\frac{\epsilon_2}{\epsilon_1}
\eeq
hence the NS limit $\epsilon_2 \rightarrow 0$ on the gauge theory side corresponds to the classical $c \rightarrow \infty$ limit in the
Liouville theory. The dimensions of the degenerate operators in the classical limit behave as 
\beq
h_{s,1}= -\frac{s-1}{2} + O(1/c) \qquad h_{1,r}= -\frac{r^2-1}{24}c + O(c^0)
\eeq

The operators are naturally classified at large $c$ limit according to  behavior of
their conformal dimensions $\Delta_i$.
The operators whose dimensions are proportional to $c$ are called heavy while ones whose
dimensions are $O(1)$ are called light operators. It is natural to introduce the
classical dimensions $\delta_i$ for the heavy operators defined as 
\beq
\Delta_i= b^{-2}\delta_i
\eeq
According to AGT correspondence the 4-point  spherical Liouville conformal block 
is identified with the instanton part $Z_{inst}$ of the total Nekrasov partition function for $SU(2)$ $N_f=4$ theory
\beq
\mathcal{F}_{a}(\epsilon_1,\epsilon_2,m_i, q) = Z_{inst}(a,\epsilon_1,\epsilon_2,m_i,\tau)
\eeq
\beq
Z_{tot}=Z_{cl}Z_{1-loop}Z_{inst}
\eeq
where the factors correspond to the classical, one-loop and instanton contribution
to the partition function. One-loop part $Z_{1-loop}$ coincides with the three-point DOZZ factor.
The coordinate at the Coulomb branch $a$ corresponds to the intermediate conformal dimension 
in the conformal block, masses $m_i$  yield the corresponding conformal dimensions 
of insertions $\delta_i$ and the complexified coupling constant $\tau = \frac{4\pi i}{g^2} + \frac{\theta}{2\pi}$
in SYM theory gets mapped
into the conformal cross-ratio in the 4-point spherical conformal block \cite{agt}.
The 4-point correlator in the Liouville theory is expressed in terms the
Nekrasov partition function as follows \cite{agt}
\beq
\lal V(0)V(\infty)V(1)V(q) \rr \propto \int da a^2|Z_{tot}(a)|^2
\eeq

In the classical NS limit
$\epsilon_2\rightarrow 0$ the twisted superpotential gets identified with the
classical conformal block .
\beq
Z_{inst}(a,\epsilon_1,\epsilon_2,\tau)\rightarrow \exp(b^{-2} W(a,m_i,\hbar,\tau))
\eeq
\beq
\mathcal{F}_{a}(\epsilon_1,\epsilon_2,\delta_i, q) \rightarrow \exp(b^{-2} f_{\delta_{in}}(\delta_i,\hbar,q)),\qquad
\delta_{in}= b^{-2}(\frac{1}{4} - \frac{a^2}{\hbar^2})
\eeq
Since we have exact coincidence of the twisted superpotential and the classical conformal block
the naive poles in the twisted superpotential correspond to the naive poles
in the intermediate dimension plane  in the classical 
Liouville conformal block. Therefore we are able to translate our findings for the superpotential the
corresponding statement that the integrand in the integral representation for the classical
Liouville correlator does not have poles.

Similarly the torus one-point classical conformal block corresponds 
to twisted superpotential in $\mathcal{N}=2^*$ theory which depends
on the external and intermediate  dimensions $\Delta,\Delta_{in}$.
To have the unified picture it is useful to represent the one-point torus conformal block $\mathcal{F}_{c,\Delta_{in}}(\lambda, q)$ as the  spherical
conformal block with four insertions $\mathcal{F}_{c,\Delta_{in}}^{sp}[\lambda_1,\lambda_2,\lambda_3,\lambda_4](q)$.
The explicit mapping of parameters under the map goes in arbitrary $\Omega$-background as follows
\beq
\mathcal{F}_{c,\Delta_{in}}(\lambda,q) = \mathcal{F}_{c,\Delta_{in}}^{sp}[\frac{\lambda}{2} - \frac{1}{2b},
\frac{\lambda}{2} + \frac{1}{2b},\frac{b}{2}, \frac{b}{2}](q)
\eeq
where conformal dimensions $\Delta_i$ are equal to
\beq
\Delta_i=\frac{1}{4}(Q^2 -\lambda_i^2) 
\eeq
The  modulus of the torus $q$ gets mapped into the position of 
the insertion $x$ on the sphere as 
\beq
q(x)= \exp \l -\frac{\pi K(1-x)}{K(x)} \r \qquad K(x)= \int_0^1 \frac{dt}{\sqrt{(1-t^2)(1-xt^2)}}
\eeq

At the classical $b\rightarrow 0$ limit the corresponding classical 4-point spherical conformal
block has two equivalent representations 
\beq
\begin{split}
f^{sp}_{\delta_{in}}[\delta_1,\delta_2,\delta_3,\delta_4](q) = f^{sp}_{\frac{1}{4} - \frac{a^2}{\hbar}}[\frac{1}{4}(1 -\frac{m^2}{\hbar}),\frac{1}{4}(1 -\frac{(m+\hbar)^2}{\hbar}),
\frac{1}{4},\frac{1}{4}](q) = \\
= f^{sp}_{\frac{1}{4} - \frac{a^2}{\hbar}}[\frac{1}{4}(\frac{3}{4}-\frac{m}{\hbar}
(1 + \frac{m}{\hbar})),\frac{1}{4}(\frac{3}{4}-\frac{m}{\hbar}
(1 + \frac{m}{\hbar})), \frac{3}{16},\frac{3}{16}](q)
\end{split}
\eeq
where $m$ is the adjoint mass in $\mathcal{N}=2^*$ theory.
 All of insertions  correspond to the heavy operators and
the intermediate classical dimension is heavy as well. 

If we insert the additional light $\Psi_{2,1}(z)$ operator in 1-point
torus conformal block and consider the 2-point classical conformal
block the Lame equation can be identified.
The decoupling equation for the 2-point block can be brought into the
conventional QM form with Lame potential
\beq
(\frac{d^2}{dz^2}+ k \wp(z|\tau) +E)\Psi(z,\tau,E)=0
\eeq
where $\tau$ is the modulus of the elliptic curve and 
$k= \frac{m}{\hbar}(\frac{m}{\hbar} +1)$
The energy in the $\wp$ potential is related to the classical 
conformal block via the properly normalized quantum Matone relation 
\beq
\frac{E}{4\pi^2}= -\l \frac{a}{\hbar} \r^2 -\frac{k}{12}(1 - 2E_2(\tau))
+\hbar^{-1}q\frac{d}{dq}W^{N=2^*}(q,a,m,\hbar) 
\eeq
where $E_2$ is the Eisenstein series and $a$ is  the Bloch phase.

The prepotential for the pure SYM theory can be derived from the 1-point torus conformal block in the 
limit that corresponds
to the dimensional transmutation in the field theory or the Inozemtsev
limit in the language of integrable systems.
Prepotential is expressed in terms of the norm
of the coherent Gaiotto state \cite{gaiotto}
\beq
\mathcal{F}_{c,\Delta}(\Lambda)=  \lal \Delta,\Lambda^2|\Delta,\Lambda^2 \rr
\eeq
\bqa
L_0|\Delta,\Lambda^2 \rr =(\Delta + \frac{\Lambda}{4}\frac{\partial}{\partial \Lambda})|\Delta,\Lambda^2 \rr,\quad \\
L_1|\Delta,\Lambda^2 \rr =\Lambda^2|\Delta,\Lambda^2 \rr ,\quad L_n|\Delta,\Lambda^2 \rr =0, n\ge2
\eqa
where $\Lambda$ is the non-perturbative scale in pure $\mathcal{N}=2$ SYM theory.

To get the Mathieu equation we  insert the probe $\Psi_{2,1}(z)$ operator 
and consider the classical limit of the 
null vector decoupling equation for the degenerate irregular conformal block. The 
wave function is represented as 
the matrix element of the degenerate chiral vertex operator between two Gaiotto 
states. 
\beq
\lal \Delta_1,\Lambda^2|V_{+}(z)|\Delta_2,\Lambda^2 \rr \rightarrow z^{\Delta_1 -\Delta_{+} -\Delta_2}
\phi(\frac{\Lambda}{\hbar},z) \exp \l \frac{1}{b^2} f_{\delta}(\frac{\Lambda}{\hbar}) \r
\eeq
The function 
\beq
\psi( \frac{\Lambda}{\hbar},z) = z^{r}\phi(\frac{\Lambda}{\hbar},z), \qquad \delta= 1/4 - r^2
\eeq
obeys the Mathieu equation with the energy 
\beq
E= 4r^2 -\Lambda \frac{\partial}{\partial \Lambda}f_{\delta}(\frac{\Lambda}{\hbar})
\eeq

\subsection{Non-perturbative QM gaps and classical  conformal block}

The phenomena we have found for the twisted superpotentials can 
be translated into the classical Liouville conformal blocks. First,
the disappearance of the poles in the twisted superpotential 
implies the absence of the poles in the intermediate dimension
in the classical conformal block multiplied by DOZZ factor.The energy 
in the QM problem is identified with the accessory parameter while 
the Bloch phase with the intermediate dimension. Recall
that accessory parameter can be derived in two ways.
First, evaluate the
classical action in the Liouville theory $S_{cl}(\delta_i,q)$ on the solution to the equation
of motion with the prescribed behavior near the points of operator insertions
\beq
\phi(z)= -2\alpha_i \log|z-z_i|+ O(1)\qquad 2b^2\Delta_i= \alpha_i(2-\alpha_i)
\eeq
Then  consider the derivative of the action with respect to the modulus
\beq
c_2=-\frac{dS_{cl}}{dq}
\eeq
which has been identified with the accessory parameter familiar in the 
uniformization problem. Generically the number of independent accessory
parameters depends on the genus of the surface and the number of the marked points.
The accessory parameter depends on the classical intermediate dimensions and function $c_2(\delta_{in})$
exactly coincides with the dependence of the energy on the Bloch phase in the QM.

The second way of evaluation of the conformal block deals with the so-called
monodromy method \cite{zam2}. One artificially inserts the additional $\Psi_{2,1}$ 
light "`probe"' operator which obeys the null-vector decoupling equation. 
Since this operator is light it does not deform the initial heavy conformal
block and indeed can be considered as the useful probe.
For instance, if we insert the 
light degenerate  $\Psi_{2,1}$ operator in 4-point conformal block the 5-point conformal block will obey 
the Lame equation with respect to the coordinate of the light operator insertion as we have discussed above. 
The poles in conformal block occur when the heavy intermediate dimension
becomes degenerate.

To define accessory parameters in the monodromy method one  utilizes the second order differential equation for the
conformal block with additional degenerate operator $\Psi_{2,1}$ inserted at point $z$.
This new 5-point conformal block obeys the second order equation  
\beq
(\frac{d^2}{dz^2} +T(z))\Psi(z)=0
\label{coad}
\eeq
with  $T(z)$ depending on the conformal dimensions of the operators
in the conformal block  and the accessory parameters $c_i$  to be determined.
The accessory parameters $c_i$ are defined via the singular behavior 
of stress tensor $T(x,z)$ near the insertions of the operators in 4-point block
\beq
T(z,x)=\sum_{i=1}^{4} (\frac{c_{i}}{z-x_i}+ \frac{h_i}{(x_i-z)^2})+ nonsingular
\eeq
They obey the sum rules
\beq
\sum_{i=1}^4 c_i =0 \quad \sum_{i=1}^4 (x_ic_i+ h_i) =0 \quad \sum_{i=1}^4(c_i x_i^2 +2c_ix_i) =0
\eeq
therefore for the 4-point block there is only one independent accessory parameter
which we denote $c_2$.
It depends on the classical intermediate dimension $\delta_{in}$  and the function 
$c_2(\delta_{in})$ derived by the monodromy method coincides with function
obtained from the derivative of classical Liouville action
with respect to the coordinate of the insertion point.

Let us summarize what kind of new information about the Liouville classical torus conformal
block can be traced from our findings concerning the partial resummation
of the instanton contributions. First, the exponentially small gap in the QM
spectrum corresponds to the exponentially small gap in the accessory parameter
$c_2(a)$ near the pole in the conformal block considered as a function of intermediate 
dimension $\delta_{in}$. The instanton series gets resummed near the pole $2a=k\hbar$ 
and trans-asymptotic matching procedure corresponds to the particular 
resummation of OPE expansion 
of the 4-point conformal block. The pole in the complex a plane at fixed $q$ 
becomes a cut in imaginary direction.
As we have shown the cut is enclosed with the
CMS in the $a$-plane and the pole disappears. 

The function $c_2(a)$ defines the
infinite genus Riemann surface for the pure $\mathcal{N}=2$ SYM case. 
For the $\mathcal{N}=2^*$ theory at special value of adjoint mass $m=\hbar$
the Riemann surface has genus one. 
To some extend the very phenomena of the cuts formation from the naive poles 
upon the proper resummation corresponds to the derivation of the Riemann surface familiar 
in the context of finite-gap solution to the KdV equations. 
It would be very interesting to understand better the meaning of the 
$\text{\bf CP}^1$ solitons in the local model from the Liouville viewpoint. We plan
discuss this issue elsewhere. Note that recently the generalization
of the recursion relations derived from the expansion of the conformal block 
in the naive poles in the exchanged momentum has been discussed in
\cite{yamazaki,gliozzi}.

It is worth
to mention one more aspect of the classical limit in the Liouville theory
namely its intimate relation with the Teichmuller space.
The classical action evaluated on the solution to the classical equation
of motion in Liouville theory plays the role of the action for the Whitham dynamics which 
on the other hand  can be interpreted as the 
geodesic motion on the Teichmuller space. The review on the relation between
the quantum Liouville theory and the quantum Teichmuller space can be found 
in \cite{teschner,teschner2}. In the context of the SW theory
relation between the prepotential and Whitham action was first noted in \cite{gkmmm,gmmm2}. 
It is important that in  the NS limit of the $\Omega$-deformed theory 
the Whitham dynamics remains classical however with perturbed Hamiltonian. The
quantization of the Teichmuller dynamics occurs only when the second parameter
of the $\Omega$-deformation is switched on.

The relation between the classical Liouville theory and 
the Whitham classical dynamics goes as follows. At the Liouville side there are
equations relating the accessory parameter $c_2$ and dual quasimomentum $a_D$
with the derivatives of classical Liouville conformal block $f(a,q)$ with respect 
to the intermediate weight $a$ ( recall that $\delta_{in}^2= \frac{1}{4} - \frac{a^2}{\hbar^2}$) 
and the insertion point $q$:
\beq
c_2= -\frac{df(a,q)}{dq} \qquad  a_D= \frac{df(a,q)}{da}
\eeq
These equations have the meaning as   conventional Hamilton-Jacobi equations
for the Whitham dynamics
\beq
E=- \frac{dS}{dt} \qquad p= \frac{dS}{dx}
\eeq
since  $(a_D,a)$ provide the Poisson pair for the Whitham phase space
and $\log q$ plays the role of time.
Additionally we have the  equation of motion in the Whitham system
\beq
\frac{da_D}{dq}= -\frac{dc_2}{da}
\eeq
It is known as P/NP relation or bridge equation in the context of QM \cite{ZJJ,dunne,pnp2} and 
its identification as the Whitham equation of motion has been found
in \cite{Whitham}. Another interpretation of this relation 
via the holomorphic anomaly and mirror transform has been developed in \cite{marino}.

The natural objects which
describe the structure of  the Teichmuller space are the coadjoint Virasoro orbits (see \cite{wittencoad,palla} for
a review)  
which are the infinite dimensional phase spaces supplemented with the 
Kirillov-Kostant symplectic form. The coadjoint
orbit is parametrized by the two-differentials $L=\frac{d^2}{dx^2} +T(x)$ where 
$T$ is stress tensor of a 2d theory. Classically
the conformal blocks are the functions on the phase space that is 
functions on the coadjoint Virasoro orbits. The classical Liouville
equation is nothing but the stationary condition for a point on the
coadjoint orbit $\dot{T}=0$ \cite{alekseev}.

Can we expect that something special can happen in the near-pole
region from the viewpoint of Virasoro coadjoint orbits? The classical
intermediate dimension at $n$-th pole corresponds
precisely to the  $\frac{Diff S^1}{SL^{n}(2,\mathbb{R})}$ coadjoint orbit.
The stabilizer of the orbit $SL^n(2,\mathbb{R})$ is formed by $(L_{-n},L_0,L_n)$ 
Virasoro generators.
Remarkably nearby this type of orbits there are two families 
of special coadjoint orbits which does not involve the constant 
representative. They can be considered as the small perturbations 
of the $\frac{Diff S^1}{SL^{n}(2,\mathbb{R})}$ orbit and are usually denoted
as $\frac{Diff S^1}{T_{n,\pm}}$, $\frac{Diff S^1}{T_{n,\rho}}$
where the stabilizers of the orbits depend on their invariants.
The presence of integer invariant $n$  reflects the fact that $\pi_1(SL(2,\mathbb{R}))=\mathbb{Z}$
and special orbits can be derived from the 
orbits of $\widehat{SL(2,\mathbb{R})}$ with nonvanishing windings by small perturbation.
Although special coadjoint Virasoro orbits  do not admit any standard quantization 
procedure they are well 
defined classical phase spaces. 

While considering classical 4-point conformal block with 
four heavy operators   four classical coadjoint
orbits are involved and the near-pole coadjoint orbit corresponds to the
intermediate dimension. We could suggest that the account of special 
coadjoint Virasoro orbits $\frac{Diff S^1}{T_{n,\pm}}$, $\frac{Diff S^1}{T_{n,\rho}}$
is important for the exponentially small 
effects in the classical  conformal blocks near poles. This point certainly
deserves the separate study.

\subsection{On non-perturbative phenomena in $AdS_3$ gravity}
\subsubsection{Mapping to the scattering in $AdS_3$}

The classical conformal blocks in Liouville theory have interesting interpretation
in the holographic picture. The central charge in the Liouville theory 
is related with  the gravitational constant in $AdS_3$ \cite{brown}
\beq
c=\frac{3 R_{AdS}}{2G_N}
\eeq
and classical limit corresponds to the small Newton constant. The spectrum
of the quantum $AdS_3$ gravity involves the vacuum state and 
energy levels corresponding to the conical defects, BTZ black holes and quasinormal modes 
in the BTZ background.

An asymptotically $AdS_3$ metrics in global coordinates reads as 
\beq
ds^2= \alpha^2 \frac{1}{\cos^2(\rho)}(\alpha^{-2} d\rho^2 +  dt^2 + \sin^2\rho d\phi^2)
\eeq
where  $\alpha^2>0$ corresponds to the conical defect while $\alpha^2<0$
to the BTZ black hole. The parameter $\alpha$ is related to the classical
dimension of the heavy boundary operator
\beq
\alpha=\sqrt{1 - 4h}
\eeq
The isometry generators of $AdS_3$ $(L_{-1},L_0,L_1)$ form the $SL(2,\mathbb{R})$ algebra and
if we introduce the variable $w=\phi +it$ the  important isometry generator
acts as $L_w= -i\partial_w$.
The temperature  of the BH is $T= 2\pi \alpha$ while  
the angle deficit equals $\delta \phi = 2\pi(1-\alpha)$ for the conical defect metric.
The threshold value $\alpha=0$ corresponds to the extremal  BTZ BH with zero temperature
or equivalently the maximally massive particle.

The precise mapping  of the  conformal blocks into the bulk has been developed
in \cite{kraus,perl1,perl2}. It was shown that the boundary conformal block
corresponds to the particular geodesic Witten diagram when the integration
over positions of trivalent vertices in the bulk goes along geodesics
connecting boundary points. If some operators are light  they correspond
to the geodesics in the $AdS_3$ background modified by the heavy operators. 
If all operators are heavy one has to take into account the equilibrium
conditions at the junctions of heavy geodesics.
Note  that previous studies have used slightly different viewpoint 
involving the Hartle-Hawking wave function in 3d gravity.
Particles correspond to the point-like insertions while the BTZ BH 
to the boundaries which  the Hartle-Hawking wave function depends on. Such viewpoint
was applied for instance for the eternal black holes \cite{maldacena} (see \cite{carlip}
for the review on this approach). 

The conformal block was identified up to a constant with the 
on-shell 3d gravity action evaluated on the bulk geodesic network \cite{Fit1,asplund,kraus}
\beq
\log\mathcal{F}(a,q,m_i)= - S_{on-shell}(a,q,m_i) = 2\sum_{segments}2h_il_i(a,q,m_i)
\eeq
where we have in mind 4-point spherical conformal block relevant for matter 
in fundamental or 1-point torus block for the adjoint. The force balance condition
is assumed at two bulk junction points in the spherical case and in one 
junction point for the torus case. The positions of vertices are subject  to the 
minimization of the total worldline action. The action of the Killing 
vector field  on the on-shell bulk action yields the conserved Killing
momentum $p_w$ which  is identified with the accessory
parameter $c_{2}$ \cite{asplund, Fit1,kraus}
\beq
q\frac{\mathcal{S}_{on-shell}}{dq}= p_{w}= -c_{2}
\eeq

Another way to reproduce the conformal block in terms of $AdS_3$ bulk 
involves the representation of $AdS_3$ gravity as $SL(2,\mathbb{R})\times SL(2,\Rbb)$ Chern-Simons
theory \cite{witten88,townsend}. In this approach the heavy operators correspond
to the flat connections with the fixed holonomies while the light operators
to the open $SL(2,\Rbb)$ Wilson lines with one end at the boundary. It was checked
in \cite{bhatta, besken} that the simplest conformal blocks are reproduced
in  the CS approach. Note that since the conformal block involving only
heavy operators concerns only flat connections such conformal blocks
can be interpreted as the specific geodesic motion on the moduli space of the flat
$SL(2,\Rbb)$ connections. The dynamics on the Teichmuller space can be obtained from 
the dynamics on  $SL(2,\Rbb)$ flat connections via Hamiltonian reduction(see
\cite{verlinde} for the early study).

In the case of heavy-light conformal blocks the bulk dynamics corresponds to the
geodesic motion in $AdS_3$ in the background created by the heavy boundary
insertions , see \cite{Fit1,Fit2,Fit3,belavin1,belavin2} for the recent studies. 
In our situation we have four heavy operators and the physical process in the bulk
is the scattering of the conical defect and the extremal BH or scattering of 
two conical defects.  We
consider the near pole regime and sum up contributions to the AdS action
from all relevant subleading saddles corresponding to SYM instantons. 
The conformal blocks involving only heavy operators have been 
previously discussed in \cite{krasnov1,krasnov2}
where the technique suitable for derivation of gravity S-matrix for the
scattering of the  particles and BH was developed. In particular the probability
of the  BH-particle scattering , Hawking emission and BH production in
particle collision has been estimated. Note that attempt to extract non-perturbative
information about 3d gravity from the semi-classical conformal blocks 
has been performed in \cite{Fit2} where the subleading saddles have 
been taken into account.

What kind of phenomena in $AdS_3$ gravity have we encountered? Is it 
non-perturbative or its interpretation in the gravity terms
does not require any tunneling? We have seen in our study that the summation of the relevant
instanton terms near the pole in the intermediate dimension in the 4-point
classical conformal block amounts
to the exponentially small gap in the accessory parameter $c_{2}$ and the higher
genus Riemann surface emerges which generically has infinite genus.

Since the accessory parameter $c_{2}$ corresponds to the conserved Killing
momenta we have therefore found the 
exponentially small gap in the values of the conserved momenta. The
intermediate dimension in the conformal block corresponds to the 
mass of the intermediate state in the scattering 
of two heavy modes and  $n$-th pole corresponds to the 
intermediate particle  with winding number $n$ in $SL(2,\Rbb)$.
Therefore we can claim that the purely winding modes in the intermediate
state in the scattering are forbidden due to "non-perturbative screening"
upon the resummation of the relevant terms in OPE.
The  AdS on-shell action develops the cuts instead of the 
naive poles in the intermediate masses 
in the $a$ plane at fixed $q$ or in the $q$ plane at fixed $a$. The latter viewpoint implies 
that transition from the early time to late time dynamics goes through the   CMS
surrounding the $q=0$ region. The rearrangement of the spectrum in $\text{\bf CP}^1$ model at the 
CMS near each naive pole in the intermediate masses corresponds to the rearrangements of  junctions of geodesics. There is
some similarity  with the geodesic representation of disappearance of the W-boson from the spectrum 
in SW theory since  the corresponding geodesics at the
$u$-plane decays into the pair of geodesics via the string junction exactly at CMS \cite{sethi}.

\subsubsection{Comparison with heavy-light conformal blocks}

Let us briefly compare our all-heavy 4-point function with heavy-light
4-point classical conformal blocks focusing at the cut structure. It was
argued in \cite{resum1,resum2} that there is the cut emanating from $z=1$
in the heavy-light conformal block $\lal O_{H}(0)O_{H}(1)O_{L}(\infty)O_L(z) \rr$
in the $z$-plane.  The presence of the
cut amounts to a few nontrivial physical phenomena emerging at small $z$ region.

The specific resummation of the small $z$ contributions has been performed in \cite{resum2}.
It was demonstrated that to reproduce the small $z$ limit at $c\rightarrow \infty $
one has to take care of the limit when $cz=constant$ in the classical limit. The
resummation of the $(cz)^n$ terms in the $z$ expansion of the conformal block has 
been performed and it was demonstrated that it reproduces correctly
the leading small $z$ power and log singularities derived in a different way. 
It is this resummation which makes the  transition from the Euclidean to Lorentzian
geometry for the conformal blocks involving the degenerate operators self-consistent.

In our case we have all-heavy classical conformal block $\lal O_{H}(0)O_{H}(1)$ $O_{H}(\infty)O_H(q) \rr$
and we consider the "`double classical limit"' although this notation could sound confusing.
Namely we consider the limit $\epsilon_2\rightarrow 0$, $\hbar \rightarrow \infty$
limit in the $\Omega$-background hence we have the product of two large numbers 
in the Liouville central charge and from the conventional classical limit viewpoint  we account
some subleading terms. We have performed the resummation of the following terms
near the pole in the intermediate dimension when the corresponding 
intermediate state becomes
degenerate 
\beq
[\frac{(\frac{q}{\hbar})^n}{n-(\frac{\delta_{in}}{\hbar})}]^k
\eeq
where $n$-is the number of pole in the conformal block when the
intermediate operator becomes $\Psi_{1,n}$.
The $\hbar$ is the large formal parameter in the classical Liouville theory 
however it effectively yields the limit of small $q$ in conformal block somewhat
similar to the heavy-light case. 

We have derived above that upon resummation the cut in the 
space-time Liouville coordinate $q$ emerges. Similar 
to the heavy-light case we have obtained the cut in the space-time
coordinate $z$ however in the all-heavy case the position of the cut
depends on the distance from the pole in the intermediate dimension.

Recall that we have started our study from the $\Omega$-deformed SYM
theory in NS limit. The Liouville space-time coordinate corresponds
to the complexified coupling constant in the SYM theory hence the cut
in the space-time Liouville coordinate near $n$-th pole corresponds to the cut in the
SYM coupling constant nearby the $n$-th vacuum state. The cut in the Liouville
theory makes the transition from the early times to the late times 
quite rich and  a kind of chaotic behavior with the Lyapunov exponents can be recognized.
Therefore we can expect the same rich structure when moving from the weak coupling
to strong coupling regimes at the SYM side of the correspondence. We hope
to discuss this point elsewhere.

\subsubsection{Analogy with the low-energy monopole scattering}

It it worth to mention the following useful  analogy. Consider the low-energy
monopole scattering at the weak coupling regime. The monopoles are heavy particles
and their scattering is described in the moduli approximation as the geodesic 
motion in the moduli space of two monopoles \cite{AH85}. The scattering phase can be evaluated
from the action along this geodesics. The moduli space of two monopoles enjoys the Atiyah--Hitchin metrics. The
non-perturbative phenomena in the monopole scattering occur due to the non-trivial geodesics
in this metrics. For instance, there is the exponentially suppressed process of the
monopoles into dyons scattering. The key physical phenomenon behind this process is that
angular momentum of the colliding monopoles gets transformed into  angular 
momentum of  electromagnetic field. It is 
angular momentum which is responsible for the 
stabilization of the Euclidean configuration responsible for this
non-perturbative process.
It can be described as the motion along a peculiar
geodesics in the moduli space which involves the motion along $S^1$ 
yielding the electric charge.
Let us emphasize that such process is possible only in some range 
of the initial angular momenta of the monopoles otherwise the relevant
geodesics are unreachable. This threshold can be also reformulated
as the existence of the threshold in the impact parameter space in the $\Rbb^3$.

On the other hand this non-perturbative process can be also
described via the exchange by light W-boson in the physical space. 
The information concerning the 
scattering phases can be extracted from the propagator of the light W-boson
which knows about its quantum spectrum. Of course this exchange can not be 
described in the naive perturbation theory since monopole is not elementary
particle and to some extend can be considered as the coherent state of the
infinite number of W-bosons.

The scattering of the heavy modes in $AdS_3$ we consider can be 
similarly described as the
geodesic motion on the Teichmuller space which can be identified with
the solution to the equation of motion in Whitham dynamics. As we have mentioned above the 
conserved Killing momentum of the colliding 
objects in 3d gravity  $c_{2}(a)$ plays the key role. Indeed, the exponentially small gap
in the values of the Killing momentum is the indication of the
non-perturbative phenomena. The emergence of the CMS in the intermediate dimension
at fixed coordinate or in coordinate at fixed intermediate dimensions in our study is the
analogue of the regions in the angular momentum plane or impact parameter plane 
when the process monopoles $\rightarrow$ dyons is possible. 

The natural 
question is if we have in the scattering in $AdS_3$ any analogue of the angular momentum transfer
from the colliding particles to the electromagnetic field similar to the monopole case. The possible speculation goes as follows. We would like to 
excite topologically non-trivial geodesics in the Teichmuller which could 
be interpreted as carrying the Killing momentum of the gravitational field.
The natural candidates for such objects are the special coadjoint Virasoro orbits
whose very existence follows from the nontrivial $\pi_1 \l SL(2,\Rbb) \r$. Hence we can 
suggest that the analog of the non-perturbative transfer of the angular 
momentum from the colliding particles to the EM field corresponds to the
"`excitation"' of the special Virasoro orbits with nontrivial 
windings along the geodesic motion
in the Teichmuller space. However note that we have found that pure winding states
in the intermediate channel disappear.  They are enclosed within the CMS where 
they decay to the solitonic states.

Similar to the monopole case we can assume that by adding the degenerate light
$\Psi_{2,1}$ probe operator, which is analogue of W-boson, in the monodromy method we probe
the "`saddle configuration  in AdS space"' responsible for the scattering
process in $AdS_3$ geometry. That is why 
the wave function of the light operator which obeys the Lame or Mathieu
equation knows about the on-shell action on the geodesic network.
The quantum spectrum of the Mathieu and Lame equations 
contains the whole information concerning the scattering process 
of the heavy objects. 

\section{Comments on  Monopole production}
\label{monopole}
In this section we will speculate on 4d interpretation of the level splitting. We conjecture that it is related to monopole
production by Omega-background. Also, in the next section we will present a 
possible brane picture.

It is instructive to take $\hbar \ra 0$
limit in order to study the original $\mathcal{N}=2\ d=4$ theory in the weak Omega-background. For large energies $u \gg \Lambda$ the gaps are exponentially small. Leading order WKB approximation
yields the following answer for the gap width \cite{Connor}:
\begin{equation}
\Delta u = \frac{\hbar}{\pi} \frac{\pr u}{\pr a} \arctan \l e^{-\frac{2\pi}{\hbar} \Im a^M_D} \r =
 \frac{\hbar}{\pi} \frac{\pr u}{\pr a} \sum_{n=1,3,5,\dots} \frac{1}{n} e^{-\frac{2\pi n}{\hbar} \Im  a^M_D}
\label{eq:wkb_gap}
\end{equation} 
For large $u$, $a_D^M$ is pure imaginary, therefore in the exponent we have the monopole mass $|a_D^M|=\Im a_D^M$.
Note that the leading term was previously recovered in eq. (\ref{mnsol}). 

Exponent $\sim \exp \l - \frac{2 \pi |a^M_D|}{\hbar} \r$ is a typical answer for a particle production rate in a weak 
external field. 
This suggests that the band structure is probably related to the Schwinger creation of
monopoles by Omega-background. This analogy becomes even more clear if we translate the energy 
(\ref{eq:wkb_gap}) into the prepotential. 
Differentiating Matone's relation (\ref{eq:TMatone}) with respect to the $a$ we obtain:
\beq
\frac{\pr a^M_D}{\pr \log \La} = -i\frac{2}{\pi} \frac{\pr u}{\pr a}
\eeq
Therefore (\ref{eq:wkb_gap}) leads to:
\beq
\label{eq:Schwinger}
\Delta F = \frac{2}{\pi} \sum_{n=1,3,5,\dots} \frac{1}{n^2} e^{-\frac{2\pi n}{\hbar} |a^M_D|}
\eeq
This is exactly the Schwinger answer for the particle production rate in a external field. 
However, there are two subtleties. First of all, if we simply integrate out monopoles in the external Omega-background
the answer is the Minkowski space\footnote{This is the reason we have $e^{i \hbar s}$ instead of $e^{\hbar s}$} 
is  given by the following integral\cite{mtop1,mtop2,Dedush}:
\beq
\label{Schwinger_int}
\frac{1}{\epsilon}\int_0^{+\infty} \ \frac{ds}{s^2} \frac{e^{-s |a^M_D|}}{e^{i s \hbar}-1}
\eeq
The integral has poles at $s= 2\pi n$. Residues at these poles produce the imaginary part which is responsible for the 
particle production. This brings us to the second subtlety: the number of pairs $n$ is always odd. We conjecture that as we jump from one band to another the imaginary part of $(\ref{Schwinger_int})$ 
changes sign for odd number of pairs. 

In the weak coupling region $u \gg 1$ the BPS spectrum contains dyons of arbitrary electric charge. 
Therefore apart from monopole production one can also expect dyon production. We conjecture that the prepotential also
receives corrections in the form
\beq
\label{conj}
\sum_{n=1}^\infty \sum_{c=0}^\infty \cfrac{(-1)^{n+1}}{n^2} \exp \l -\frac{2 \pi n (|a^M_D| + c a)}{\hbar} \r
\eeq
The above equation surely received $\hbar$ corrections. It is natural to guess that these corrections are purely perturbative,
in other words they can be easily calculated by substituting $a,a_D$ by the exact WKB periods in the spirit of 
\cite{Morozov_WKB}. It would be interesting to verify all these conjectures by comparing eq. (\ref{conj}) with exact 
WKB analysis similar to that of Kashani-Poor and Troost\cite{Troost}.

\subsection{Brane picture}
As we have demonstrated, the Nekrasov partition function in the 
NS limit $\eps_2 \ra 0$ has branch cuts at $2a=n \hbar = n \eps_1, n \in \mathbb{Z}$. 
In the effective 2d description these poles
are related to W-boson decay. It is interesting to find the interpretation of these poles from the 
viewpoint of the original 4d theory. WKB result (\ref{eq:Schwinger}) suggests that
the cuts are related to monopole production by the Omega-background, since the jump at the cut is proportional to monopole production rate. 
But why do monopoles appear only then $2a$ is an integer multiplier of $\eps_1$? 
Unfortunately, we do not have a clear explanation for the \textit{NS limit} $\eps_2 =0$. However, as we will 
demonstrate in this section, 
in the so-called \textit{unrefined limit} $\eps_1=-\eps_2=\eps$ one can easily connect the 
condition $2a=n \eps_1$ to monopole production even in the $SU(N)$ case.

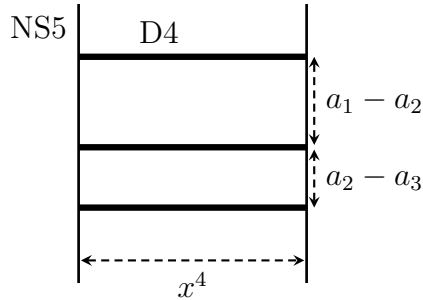
\begin{figure}
\centering
\begin{tikzpicture}[brane/.style={line width=1pt}, d4/.style={line width=2.4pt}]
  \draw[brane] (0,0) -- (0,3.7) node [below left] {NS5};
  \draw[brane] (3,0) -- (3,3.7);
  \draw[d4] (0,1) -- (3,1);
  \draw[d4] (0,1.8) -- (3,1.8);
  \draw[d4] (0,3) -- (3,3) node [midway, above left] {D4};
  \draw[densely dashed,{stealth}-{stealth},thick] (0.03,.3) -- (2.97,.3) node [midway, below] {$x^4$};
  \draw[densely dashed,{stealth}-{stealth},thick] (3.1,1) -- (3.1,1.77) node [midway, right] {$a_2-a_3$};
  \draw[densely dashed,{stealth}-{stealth},thick] (3.1,1.83) -- (3.1,3) node [midway, right] {$a_1-a_2$};
\end{tikzpicture}
\caption{$U(3)$ pure gauge theory. The distance $x^4=1/g^2_{\text{4d}}$ between NS5 branes in the 4-direction is the 
instanton counting parameter.
Coordinates of D4 branes in the $(x^5,x^6)$ plane set the corresponding Higgs VEV.}
\label{fig:d4_un}
\end{figure}

\begin{figure}[h!]
\centering
\includegraphics{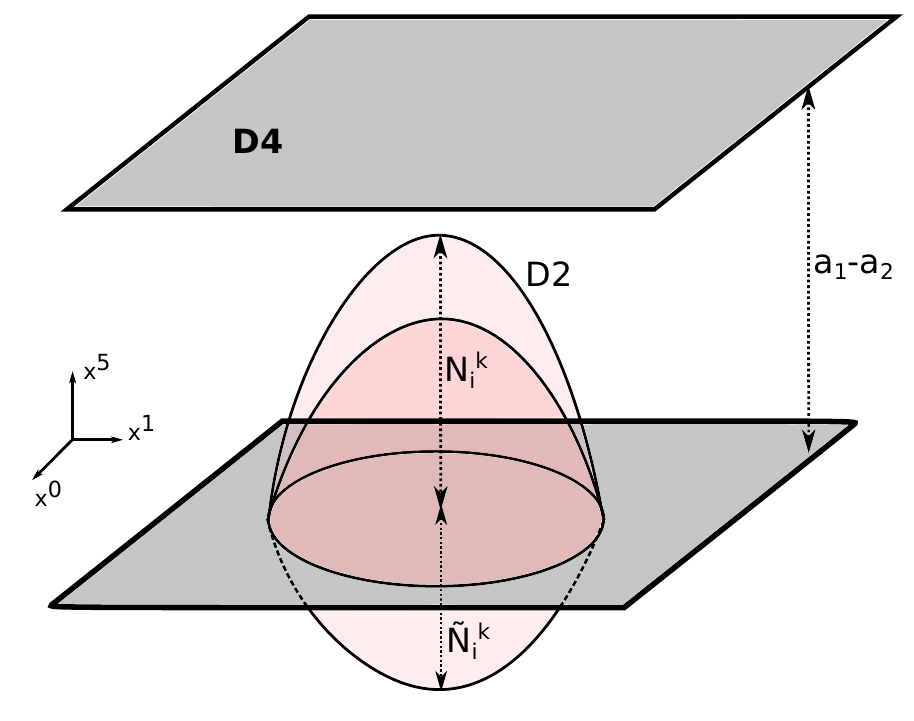}
\caption{Geometry of composite D0 -- fuzzy D2-$\overline{\text{D2}}$ bubble. Distance $N_i^k = \eps \l n_i^k + 1/2 \r$. 
For each D4 brane at $x^5=a_i$, there could be an arbitrary number of bubbles extending to $a_i+\eps \l n^k_i + 1/2 \r$
 on one side and
$a_i-\eps \l \tilde{n}^j_i + 1/2  \r$ on the other. The only requirement is that D2 should not coincide: $n^k_i \textless n^{k+1}_i $}
\label{fig:bubble}
\end{figure}

We start from type IIA brane construction for pure $\mathcal{N}=2$ $SU(N)$ gauge theory without Omega-background.
We suspend $N$ D4 branes between two NS5 branes (see Figure \ref{fig:d4_un}).
Branes are stretched along the following directions:
\begin{center}
\begin{tabular}{rl}\toprule
  NS5: & 0 1 2 3 \phantom{4} 5 6 \\
  D4: & 0 1 2 3 4 \\\bottomrule
\end{tabular}
\end{center}
Positions of $N$ D4 branes in the $(x^5,x^6)$ plane determine the Higgs VEV $a_i$.
Instantons are represented by D0 branes stretched along $x^4$ between two NS5. W-bosons are realized by fundamental strings between D4. 
Monopoles are D2 branes stretched along $(x^0,x^4,l)$, where $l$ is a line in the $(x^5,x^6)$ plane connecting two D4 branes.
In \cite{Matsuura} it was demonstrated that in this construction, the Omega-background corresponds to 
the constant flux of Ramond-Ramond(RR) 4-form field strength:
\beq
F^{(4)}=2 \eps \l dx^4 \wedge dx^5 \wedge dx^1 \wedge dx^0 - dx^4 \wedge dx^5 \wedge dx^2 \wedge dx^3 \r
\eeq
In this background D0 branes turn into a bound state of D0 and fuzzy D2-$\overline{\text{D2}}$ due to the Myers effect - Figure \ref{fig:bubble}. Moreover, one can recover the Nekrasov partition function\cite{Matsuura}:
in the $(x^5,x^6)$ plane D2 branes effectively interact as point-like 2d Coulomb charges, with +1 charge at $a_i+\eps n^k_i$ and -1 at $a_i+\eps \tilde{n}^k_i$. The potential between a charge $q$ at $x$
and charge $q'$ at $y$ is of course logarithmic:
\beq
V = -q q' \log |x-y|
\label{eq:pot}
\eeq 
The partition function of this 2d Coulomb gas directly leads to the Nekrasov partiton function. 
In fact, there is a one-to-one correspondence between integer numbers $n^k_i, \tilde{n}^k_i$ and a Young diagram
$\la_i$.

Now lets return to the monopole production. Note that the D0-D2 composites, like the original D0 brane instantons, are stretched along the $x^4$ direction. The only difference between them and monopoles
is that monopole D2 branes are stretched all the way from one D4 to another.
We conjecture that if the distance between two of the D4 branes becomes an integer multiple of $\eps$, then D0-D2 may form a cylindrical D2.
But cylindrical D2 brane is exactly the Euclidean configuration responsible for monopole production\cite{Gor2001}!
The formation may occur in two situations: either two D2 branes coming from different D4 touch and form a cylinder or 
one D2 becomes big enough to overlap the adjacent D4. 
In this case we have a cylinder plus an instanton starting from the adjacent D4. It is easy to see that such collisions indeed lead to poles in the Nekrasov partition
function: contribution to the partition function from two D2 branes is proportional to($k,l$ simply numbers D2 branes)\cite{Matsuura}:
\beq
e^{-V} = \frac{1}{a_i + \eps  n_i^l  - a_j + \eps \tilde{n}_j^k + \eps}
\eeq
Therefore if two D2 touch, it indeed leads to a pole. Apart from the potential energy (\ref{eq:pot}) each D2 brane has a kinetic energy:
\beq
e^{K_i} = \prod_{l=1}^N \frac{1}{\l a_l-a_i - \eps \tilde{n}_i^l \r \l a_l-a_i - \eps \l \tilde{n}_i^l - 1  \r \r \dots \l a_l-a_i + \eps n_i^l \r}
\eeq
Therefore if D2 brane intersects adjacent D4, it leads to a pole too.

To further check this picture lets study a single D2 brane cap stretched between two D4 in $SU(2)$ theory. The corresponding
pair of
Young diagrams is just a row with $n$ boxes if $2a = \eps n$ and an empty diagram.
We are expecting that such configuration becomes a single cylinder stretched between two D4 and its 
contribution should be of order $e^{-2\pi |a_D^M|/\eps}$ for large $n$. Indeed, it is easy to check
that for such pair of Young diagram the
Nekrasov formulas produce
\beq
\frac{1}{n!^2 (n+1)!^2} \sim \exp  \l -\frac{4 \pi |a^M_D|}{\eps} \r
\eeq
where we have used eq. (\ref{adm}). Note the appearance of the same factor\footnote{The 
difference $n \ra n-1$ appears because here we are dealing with self-dual Omega-background.}  
$\frac{1}{n!^2(n+1)!^2}$ as in eq. (\ref{eq:gn}). We expect that after the resummation of an infinite number of 
such cylinders we will again obtain the square root function as in (\ref{eq:gn}), such that the contribution of 
a single cylinder in indeed $e^{-2\pi |a^M_D|/\eps}$.

\section{Conclusions}

In this paper we have analyzed the interplay between the non-perturbative tunneling effects in the
QM and the fate of the naive singularities in the effective twisted superpotential of the effective two-dimensional theory.
We have focused on the simplest phenomena in this context - exponentially small gaps in
the QM  spectrum well above the barrier.
It was shown explicitly that 
each pole singularity gets split into the cut upon the summation over the proper terms in the instanton 
trans-series. This procedure
corresponds to the trans-asymptotic matching procedure known in mathematical literature. 
In the case of pure $U(2)$ gauge theory the theory that appears in the local description was identified with the sigma model on $\mathbf{CP}^1$  and two SUSY  vacua in 2d theory 
support the existence of the solitonic 2d states. Much similar to the decay
of the $W$-boson in the SW theory, the $W$-boson with angular momentum $n$ decays into the pair
of solitons inside the curve of marginal stability around $n$-th cut.
Some partial results have been also obtained for $U(N)$ case where the corresponding local model was identified with
the flag sigma-model.


The non-perturbative twisted superpotential for $SU(2)$ with fundamental or adjoint matter theory is known due to AGT correspondence to coincide with the classical 
conformal block. The naive pole in the superpotential corresponds to the naive pole in the classical conformal block where 
the intermediate dimension corresponds to the  degenerate operator. We argue that near each pole the conformal block enjoys the $\mathbf{CP}^1$ 
geometry. Moreover there is a wall crossing phenomena near each pole which can thought of as the region in the space of intermediate dimensions or in the complex time plane near 
the operator insertion point. Using the holographic correspondence we can map near the pole geometry of the classical conformal block into the scattering of two 
heavy objects in $AdS_3$ geometry. We have argued that
the poles in the scattering amplitudes corresponding to the pure winding modes in the intermediate states disappear
and the cuts get developed. We predict some wall-crossing phenomena for the scattering process of conical defects and extremal black holes. However this issue 
deserves much more detailed study. 
The analogy with the non-perturbative effects in the low-energy monopole scattering involving the geodesic motion on the monopoles moduli space has been mentioned.

It is a bit surprising in spite of the resurgence arguments that the exponentially 
small gap in the spectrum due to the above barrier reflection  which certainly
is the effect of instanton--anti-instanton pairs in QM is expressed in terms of the instantons
in 4d without the use of anti-instantons. To get more intuition on this subtle point we have taken
a look at the Schwinger-like interpretation of the exponentially gap formation in 4d theory as some 
instanton--anti-instanton phenomena as well. There are arguments
that the gap formation is related to Schwinger non-perturbative monopole pair 
creation by the graviphoton. However this point certainly deserves further analysis. Note that
there is some analogy with the 4d SYM case when the effects of monopoles in 4d can be recovered
upon the summation over the instantons in (4+1)d theory with one small compact dimension \cite{lawrence}.
In our case we have recovered the effects of 2d solitons summing up the monopole loops which correspond
to the bounces for the pair creation and lives effectively in 2+1 dimensions.

There are many questions to be answered. The difficult question concerns the
interpolation between large and small Planck constant in QM. The regions in the moduli space around each cut
enclosed by the CMS at large $\hbar$  have to be glued into the single CMS around the origin of the
moduli space in SW solution at $\hbar=0$. This implies the complicated rearrangement of the wall-crossing 
network when $\hbar$ is decreasing. The complexification of the Planck constant 
discussed in \cite{marino2} as well as approach suggested in \cite{couso} seem to be 
important for this question. The analysis of 
the band and gap structure 
in the full $\Omega$-background certainly of great interest. It corresponds to the 
quantization of the Whitham dynamics and being translated via AGT to Liouville theory
and $AdS_3$ gravity concerns the non-perturbative effects in quantum 3d gravity which
would generalize our consideration of the semi-classical scattering. 

The generalization
of the analysis for the 5d SYM gauge theory and therefore for the relativistic Toda system 
seems to be straightforward and the interplay between the summation over the instantons
near the naive poles and summation over the torus knot invariants \cite{gms} 
could yield new ways to utilize the knot invariants in context of the tunneling phenomena 
in the quantum mechanics with difference Schr\"odinger equation. The 
interplay between the naive poles and non-perturbative completion 
of the theory has been discussed in ABJM theory in \cite{abjm1,abjm2}. It 
would be interesting to apply our approach to that case. Another 
interesting question concerns the possible meaning of our analysis
in the hydrodynamical picture for the instanton moduli space at large $N$ \cite{hydro1}.

The list of stable BPS objects in $\Omega$-background 
involves the BPS domain walls and BPS strings \cite{otherbps1,otherbps2,bps}
which saturate the corresponding central charges in SUSY algebra.
It is not clear which role these 4d states play in the sigma models 
near the cuts on the QM side and in the $AdS_3$ gravity. In particular the most pressing 
question concerns the interpretation of the 2d kinks in the
near-pole $\mathbf{CP}^1$ model from the 4d viewpoint. In the worldsheet theory on the
surface operator or on the non-Abelian string the kinks correspond
to the trapped monopoles \cite{tong}. However in our case when 
we have just the dimensional reduction to 2d in the $\Omega$-background
such interpretation does not work literally or at least needs some modification.
Another question is: what is the counterpart of BPS domain walls and strings on the QM side?
The recent development in \cite{bullimore} seems to be relevant for this question.

One more question concerns the issue of the "time decay of the correlators"
which becomes popular in the context of the information loss paradox starting 
from \cite{shenker}.
In the Liouville context one treats the coordinate of the vertex insertion
as time-like variable and investigate the transition from the early time to
late time behavior of the conformal block focusing at the subleading saddles 
which yield non-decaying late-time contributions. In our study we have 
found that there are wall-crossing phenomena near the naive poles. Hence the transition 
from early time to late time behavior could undergo the wall-crossing at intermediate times
which is purely non-perturbative phenomena.

\section*{Acknowledgments}
We would like to thank G.~Basar, K.~Bulycheva, O.~Costin, G.~Dunne, F.~Popov, E.~Gorsky,
J.~Kaplan, P.~Koroteev, S.~Lukyanov, N.~Nekrasov, G. Tarnopolsky for 
useful discussions. A.G and N.S. are grateful KITP at UCSB for the  
hospitality during KITP program “Resurgent Asymptotics in Physics and
Mathematics"' where the part of the work has been done. A.G. thanks 
Basis Foundation fellowship for the support. The work of
A.G. was performed at the Institute for Information Transmission Problems with the financial support of the Russian Science Foundation (Grant No.14-50-00150).
N.S. acknowledges support of IHÉS during his visit in 2017 and
funding from the European Research Council (ERC) under the European
Union’s Horizon 2020 research and innovation program (QUASIFT grant
agreement 677368).

\appendix
\addcontentsline{toc}{section}{Appendices}
\section*{Appendices}
\section{g-functions for pure $U(2)$}
\label{sec:gs}
Here we list a few $g^{(n)}_k$ functions which have appeared in the main text for pure $U(2)$ theory. 

\begin{itemize}
\item $n=1$
\bqa
g^{(1)}_1(z) = -\frac{\sqrt{4 z^2+1}-\log \left(\frac{1}{2} \left(\sqrt{4 z^2+1}+1\right)\right)-1}{z}\\
g^{(1)}_2(z) = -\frac{\left(2-z^2\right) \sqrt{4 z^2+1}-\left(3 z^2+2\right)}{12 z^3}\\
g^{(1)}_3(z) = -\frac{\frac{127 z^4}{288}-\frac{z^2}{16}+\frac{\frac{11 z^6}{180}-\frac{599 z^4}{1440}+\frac{13 z^2}{80}+\frac{1}{20}}{\sqrt{4 z^2+1}}-\frac{1}{20}}{z^5}\\
g^{(1)}_4(z) = -\frac{-\frac{16985 z^6}{20736}+\frac{1847 z^4}{5184}-\frac{z^2}{64}+\frac{\frac{55 z^{10}}{1008}+\frac{235 z^8}{567}-\frac{191617 z^6}{145152}-\frac{4343 z^4}{36288}+\frac{71 z^2}{448}+\frac{1}{42}}{\left(4 z^2+1\right)^{3/2}}-\frac{1}{42}}{z^7}
\eqa

\item
$n=2$
\bqa
g^{(2)}_1(z) = -\frac{\sqrt{z^2+1}-\log \left(\frac{1}{2} \left(\sqrt{z^2+1}+1\right)\right)-1}{z}\\
g^{(2)}_2(z) = -\frac{8-8 \sqrt{z^2+1}}{9 z}\\
g^{(2)}_3(z) = -\frac{-\frac{3355 z^2}{1728}-\frac{-\frac{11141 z^4}{5184}-\frac{3679 z^2}{1728}-\frac{3}{8}}{\sqrt{z^2+1}}-\frac{3}{8}}{z^3}
\eqa

\item
$n=3$
\bqa
g^{(3)}_1(z) = -\frac{\sqrt{\frac{z^2}{36}+1}-\log \left(\frac{1}{2} \left(\sqrt{\frac{z^2}{36}+1}+1\right)\right)-1}{z}\\
g^{(3)}_2(z) = -\frac{3 z^2-3 z^2 \sqrt{\frac{z^2}{36}+1}}{16 z^3}
\eqa

\end{itemize}

\section {Comment on two Riemann surfaces}
The Lame equation  at fixed coupling is a simple example of the finite-gap potential familiar in the
theory  of the KdV equation. In general the potential $n(n+1)\wp(x)$ corresponds to genus $n$ Riemann surface and
therefore the Mathieu cosine potential which can be derived from the Weierstrass function via the Inozemtsev 
limit $n \rightarrow \infty$ corresponds to the infinite genus Riemann surface.

It is worth to make a comment on the place of these Riemann surfaces in our study. First of all recall that there are two 
Riemann surfaces in the play. The first one $P(x,y)=0$  appears at the classical level and corresponds to the fixed level
of all integrals of motion $I_k$  in the holomorphic Hamiltonian system $I_k(p_i,q_i)=I_k$. For $SU(2)$ case this is genus one Riemann surface bundled
over the complex energy plane. For $SU(N)$ it is higher genus Riemann
surface, for instance $g=N-1$ for pure $SU(N)$ gauge theory.  Upon quantization this Riemann surface 
becomes the operator yielding the Baxter equation acting on the "`wave function"' of the single separated
variable $x$:	
\beq
P \l x,\frac{d}{dx} \r Q(x,I_k)=0
\eeq
The commutation relation between $x$ and $y$ can be more complicated and the Baxter equation can be 
a difference or even an integral equation. The $x$-dependence of wave function of the quantum integrable 
system $\Psi(x,E)$ at fixed energy or fixed integrals of motion in the many-body case
is controlled by the solutions to the Baxter equation.

However there  is the second Riemann surface  which governs the $E$-dependence of the wave function
and its genus is not related with the genus of the  classical spectral curve. The potential in  Schr\"odinger equation
is treated as the initial condition for the KdV evolution and the whole KdV dynamics occurs 
on this finite genus Riemann surface. Let us remind how it can be obtained.  Consider the
Schr\"odinger equation with the periodic potential
\beq
(-\frac{d^2}{dx^2} +V(x)- E) f(x)=0
\eeq
Two solutions of the Schr\"odinger equation defines the monodromy matrix $M$
\beq
 \l f_1(x+T,E),f_2(x+T,E) \r= \l f_1(x,E),f_2(x,E) \r M
\eeq
The Riemann surface can be written in terms of the monodromy matrix  in the form
\beq
t^2- \Tr M t +1 =0
\eeq
and relates the complex Bloch phase of the solution and its energy. 

For the potential $V=-2 \wp(x)$ the genus one Riemann surface reads as
\beq
y^2=\prod_{k=1}^{3}(E-E_k)
\eeq
where
\beq
E_1=\wp(\omega_1),\qquad E_2=\wp(\omega_2),\qquad E_3=\wp(\omega_3), 
\eeq
and $(2\omega_1, 2\omega_3)$ are fundamental periods of $\wp$ function,$\sum \omega_i =0$.
 For the adjoint mass in the $\nsusy=2^*$ theory $m_{ad}=\hbar n$ 
we have $2n$ naive poles in the superpotential which yield the $n$ cuts upon the split. For
pure $\mathcal{N}=2$ SYM we have the infinite number of poles and this fits with the fact the Riemann
surface for the Mathieu potential has infinite genus.

\printbibliography

\end{document}